\documentclass[nofootinbib,tightenlines,superscriptaddress,showpacs]{revtex4-2}
\usepackage{amssymb,amsmath,amsthm,graphicx,subfigure,float,cancel,braket,comment,tipa,bm,dsfont,graphicx,hyperref}
\usepackage{xcolor}
\usepackage{mathrsfs,enumitem,newtxtext} 
\hypersetup{
	colorlinks,
	linkcolor={red!50!black},
	citecolor={blue!50!black},
	urlcolor={blue!50!black}
}
\usepackage{newtxmath}
\usepackage[T1]{fontenc}

\begin{document}
	\title{Supersymmetric higher-derivative models in quantum cosmology}
	\author{Nephtalí Eliceo Martínez Pérez}
	\email{nephtali.martinezper@alumno.buap.mx}
	\author{Cupatitzio Ramírez Romero}
	\email{cramirez@fcfm.buap.mx}
	\affiliation{Facultad de Ciencias Físico Matemáticas, Benemérita Universidad Autónoma de Puebla}

\begin{abstract}
	We study the quantum cosmology of supersymmetric, homogeneous and isotropic, higher derivative models. We recall superfield actions obtained in previous works and give classically equivalent actions leading to second order equations for the bosons, and first order for the fermions. Upon quantization, the algebra of fermions leads to a multi-component state, which is annihilated by the Hamiltonian and supersymmetric constraint operators. We obtain asymptotic wave functions of the oscillatory type, whose classical limit corresponds to inflationary evolution, and exact exponential wave functions. We use the latter to derive probability distributions of the initial curvature that are compatible with those obtained using the non-supersymmetric model.
\end{abstract}

\maketitle
\section{Introduction}\label{sec1}
On the largest scales, the Universe is well described by the Friedmann equations for a spatially homogeneous and isotropic universe with a perfect fluid. Evolving backwards in time, one approaches an initial singularity, a state of infinite curvature and energy density at the origin of the universe \cite{halliwell}. Most likely, the singularity signals the breakdown of general relativity, a classical theory, and rather quantum gravity should be considered \cite{Kiefer}.

Quantum cosmology aims at shedding light on the origin and subsequent history of the Universe by invoking ideas of quantum theory. On the one hand, we need a dynamical law, that is, a theory of quantum gravity. Traditional quantum cosmology is based on the canonical quantization of general relativity or quantum geometrodynamics. The fundamental equation, called the Wheeler-DeWitt (WDW) equation, is defined on \textit{superspace}, the configuration space of all 3-metrics (modulo diffeomorphisms) and matter configurations on a spacelike hypersurface. On the other hand, boundary conditions fixing the state at the boundary of superspace must be prescribed \cite{halliwell,wiltshire,hartle}. The output is the wave function of the Universe, depending only on the spatial metric and the matter fields. At this point, one still has to face fundamental issues such as the Problem of Time \cite{kuchar,ramirez2022}.

In solving the WDW equation, one usually makes some additional simplifying assumptions \cite{hawking84}. When it comes to cosmology, spatial homogeneity and isotropy constitute a reasonable first step towards a quantum description of the Universe \cite{Socorro_2021}. Such symmetries imply a drastic simplification, from the infinite dimensional superspace to the finite dimensional \textit{mini-superspace}, consisting of the scale factor and a few homogeneous matter fields. This leads to the study of constrained quantum mechanical systems. In this work we study mini-superspace Friedmann-Lema\^itre-Robertson-Walker (FLRW) models with quadratic curvature and supersymmetry.

Quadratic curvature terms were incorporated into the effective gravitational action as loop corrections arising from quantized matter fields \cite{birrelldavies}. Moreover, the necessity of higher order curvature terms, or rather higher derivatives, had been recognized to alleviate UV divergences in perturbative quantum gravity \cite{stelle}. Quadratic curvature is also motivated by the inflationary model of Starobinsky \cite{starobinsky1980}, which has made it through the ever more detailed cosmological observations and is still in good agreement with the latest constraints \cite{martin,Chaichian}. In fact, most single field slow-roll models share some resemblance with the Starobinsky model \cite{ketovn,Ketov_2020}.

Supersymmetry (SUSY), on the other hand, is a graded symmetry that unifies non-trivially internal and spacetime symmetries \cite{wessbagger}. That it has not yet been detected can be explained by the fact that supersymmetry must necessarily be broken, and the energy of SUSY breaking is well above the range of exploration of current particle accelerators. Fortunately, there is another prospect to prove supersymmetry. If SUSY is restored at higher energies, it could well have played a major role in the early universe and left an imprint in the primordial cosmological perturbations \cite{baumanngreen}.

As is well known, a theory of local SUSY, called generically supergravity, includes general relativity. Since we are interested in a cosmological scenario, we shall consider some form of FLRW supergravity. That supergravity provides a sort of square root for gravity \cite{macias} has a direct application in quantum cosmology. Upon quantization, we get additional equations that in some cases lead to unique wave functions, constituting in this way a form of boundary conditions provided automatically by the theory.

There are many approaches to supersymmetric cosmology which, according to the starting point, can be divided into two main lines. On the one hand, a consistent reduction of the degrees of freedom of N=1 4D supergravity while keeping spatial homogeneity and isotropy leads to a theory with N=4 local SUSY \cite{moniz}. On the other hand, adding fermionic degrees of freedom to the reduced FLRW action. This can also be done in several ways. For example, given a Hamiltonian and a solution of the Euclidean Hamilton-Jacobi equation, a pair of fermionic supercharge operators can be defined in such a way as to recover N=2 local SUSY \cite{Lidsey2000}.

In this work, we use the superfield formulation of 1D supergravity detailed in \cite{ramirez2008,garcia,tesis} (cf. \cite{Holten2018}), constructed on the earlier works \cite{obregon1,obregon2}. This formalism allows to write Lagrangians with time-dependent supersymmetry. We use an FLRW curvature superfield $\mathcal{R}$ which, as in 4D supergravity \cite{wessbagger}, does not follow from geometric considerations but is ad-hoc. It must be emphasized that the possible equivalence of the different approaches to supersymmetric cosmology is still an open issue.

The objective of this work is to study higher derivative FLRW cosmological models, with emphasis on the model of Starobinsky, in the context of quantum cosmology. For that purpose, we first recall and further develop the models in references \cite{nephtali21,tesis}. More specifically, we give a different second order formulation of the higher derivative Lagrangians, such that, in the Hamiltonian formulation, the supersymmetric constraints are linear in the bosonic momenta, instead of quadratic as in the previous works. This, in turn, allows a quantization where the quantum state has less components and the physical content of the theory can be appreciated more clearly. Also, some of the models are generalized by including a general $F(\mathcal{R})$ function in the Lagrangian and positive spatial curvature.

The paper is organized as follows. In Section \ref{sec2}, we briefly review the quantum cosmology of the non-supersymmetric flat model of Starobinsky. In Section {\ref{sec3}} we describe an N=1 supersymmetric $k=0$ quadratic curvature model. We obtain approximate wave functions and compare them with those of the non-supersymmetric case. Section \ref{quadratic} has to do with the N=2 models. For completeness, we briefly describe in Section \ref{secfdr} the quantization of $F(\mathcal{R})$ models. Classically, they lead to second order fermionic equations of motion and, in this sense, they sit in between the linear model $F(\mathcal{R})=\mathcal{R}$ \cite{ramirez2016}, leading to first order equations, and the more general models derived from actions of the form $(\nabla \mathcal{R})^2-F(\mathcal{R})$ (where $\nabla$ stands for the covariant derivative), leading to third order equations. The latter kind of models are the topic of Section \ref{arbi}. We focus on a model whose bosonic part is Starobinsky and a massive scalar field. We discuss the classical limit and provide approximate and exact solutions of the quantum supersymmetric constraints. Further, we give the wave functions for arbitrary $F$. As an application, we compute the probability distribution of the initial value of the bosonic curvature, which is interpreted as the wave function predicting or not an inflationary universe. Lastly, in Section \ref{sec5} we draw some conclusions.
 
\section{The FLRW model of Starobinsky}\label{sec2}
FLRW universes possess a system of local coordinates $\{t,r,\theta,\varphi\}$ in which the line element takes the form $ds^2=-N^2(t) dt^2+a^2(t) [(1-k r^2)^{-1} dr^2+r^2 (d\theta^2+\sin^2 \theta d\varphi^2)]$. The only undetermined functions are the dimensionless scale factor $a(t)$ and the lapse $N(t)$. At a given time, the curvature of the spatial sections is $^3 R=2 k a^{-2}$ where $k$ can be either positive, negative or zero \cite{ellis}. In this paper, we are interested in the model of Starobinsky,
$S=\frac{1}{16 \pi G} \int d^4 x \sqrt{-g} \big(R+\frac{\alpha}{6} R^2\big)$, where $G$ is Newton's constant $(c=1)$. In this case the scalar 4-curvature is given by
\begin{align}\label{frwcurvature}
	R_{\text{FLRW}}=6 \left( \frac{1}{aN} \frac{d}{dt} \big(\frac{\dot{a}}{N}\big)+\frac{\dot{a}^2}{a^2N^2}+\frac{k}{a^{2}}\right) .
\end{align}
and $\sqrt{-g}=N a^3 (1-k r^2)^{-1/2} r^2 \sin \theta$. Performing the spatial integral of $S$ over a region of finite co-moving volume $\mathcal{V}_0$ yields 
\begin{align}\label{starlag}
S=\frac{3}{\kappa} \int dt N a^3 \left( \frac{1}{6} R_{\text{FLRW}}+\frac{\alpha}{36} R_{\text{FLRW}}^2\right) ,
\end{align}
where $\kappa=8 \pi G/\mathcal{V}_0$. From now on the label FLRW will be omitted.

The Hamiltonian formulation of the higher-derivative action (\ref{starlag}) can be obtained with the Ostrogradsky method \cite{nephtali21,woodard}. Alternatively, one can rewrite the theory treating a combination of fields as independent. Following \cite{hawkingluttrell}, we introduce the new variable
\begin{equation}\label{phi}
	\phi=a \big(1+\frac{\alpha}{3} R\big)
\end{equation}
For nonzero $\alpha$, the limit $\phi=a$ corresponds to vanishing $R$.

In terms of the variables $\{N, a, \phi\}$ the Lagrangian and Hamiltonian of the Starobinsky model (\ref{starlag}) are
\begin{align}
	L&=\frac{3}{\kappa} \Big[-a \frac{\dot{a} \dot{\phi}}{N}+k N \phi-\frac{1}{4 \alpha} N a (\phi-a)^2\Big] \label{staraction}\\
	H&=N H_0 \equiv N \Big[-\frac{\kappa}{3} a^{-1} p_a p_\phi-\frac{3}{\kappa} k \phi+\frac{3}{4\kappa \alpha} a (\phi-a)^2 \Big]\approx 0. \label{Hamiltonian}
\end{align}
The vanishing of the Hamiltonian follows from consistency of the primary constraint $p_N=0$. By using light-cone coordinates $x=\frac{1}{\sqrt{2}} (\phi+a)$, $y=\frac{1}{\sqrt{2}}(\phi-a)$, the kinetic term can be diagonalized, $-p_y^2+p_x^2$ \cite{hawkingluttrell}. The negative kinetic energy is characteristic of the scale factor, while the scalaron $f'(R)$ has positive energy.

In terms of coordinates and velocities, the Hamiltonian equations of motion, $\dot{p}_a=-\partial H/\partial a$, $\dot{p}_\phi=-\partial H/\partial \phi$, and the Hamiltonian constraint $H_0\approx 0$ read, respectively
\begin{subequations}
\begin{align}
	\ddot{\phi}+\alpha^{-1} \phi-\frac{3}{4\alpha} a-\frac{1}{4 \alpha} \frac{\phi^2}{a}=0, \label{eqf2}\\
	\frac{\ddot{a}}{a}+\frac{\dot{a}^2}{a^2}+\frac{k}{a^2}+\frac{1}{2\alpha} (1-\frac{\phi}{a})=0 \label{eqf1}, \\
	\frac{\dot{a}}{a} \dot{\phi}+\frac{k}{a^2} \phi-\frac{1}{4\alpha} a \big(\frac{\phi}{a}-1\big)^2=0. \label{eqf3}
\end{align}
\end{subequations}
For $\dot{a}\ne 0$, (\ref{eqf3}) is a first integral of (\ref{eqf2}). On the other hand, (\ref{eqf1}) gives the on-shell value of $\phi$ (\ref{phi}), whose substitution into (\ref{eqf3}) yields the second-order Friedmann equation
\begin{align}
	\ddot{H}-\frac{\dot{H}^{2}}{2H}+3H \dot{H}+\frac{H}{2\alpha}=-\frac{k}{a^2} H \left(\frac{k}{2 a^2 H^2}+\frac{1}{2\alpha H^2}-1\right)  \label{friedmannstar},
\end{align}	
where $H\equiv \frac{\dot{a}}{a}$ is the Hubble factor. Finally, adding equations (\ref{eqf2}) and (\ref{eqf3}), and using (\ref{eqf1}) to eliminate any explicit dependence on $k$, we get $\ddot{\phi}+\frac{\dot{a}}{a} \dot{\phi}+\alpha^{-1} (\phi-a)-\phi (\frac{\ddot{a}}{a}+\frac{\dot{a}^2}{a^2})=0$, which is equivalent to
\begin{align}
	\ddot{R}+3 H \dot{R}+\alpha^{-1} R=0,
\end{align}
showing that the 4-curvature behaves as a scalar field of mass $M=\alpha^{-1/2}$.

Now we look for an inflationary instance of the equations of motion for $k=0$. During slow-roll inflation \cite{baumann}, $|\epsilon| \equiv |-\frac{\dot{H}}{H^2}| \ll 1$. Requiring also $|\frac{1}{H} \frac{d\epsilon}{dt}|\ll 1$ we get $=|\frac{\ddot{H}}{H^3}|\ll 1$. Under these conditions the relevant equations are
\begin{subequations}
	\begin{align}
		\dot{H}&\approx -\frac{1}{6 \alpha} \label{hinf}, \\
		R&\approx 12 H^2, \label{RH} \\
		\dot{R}& \approx -\frac{4}{\alpha} \big(\frac{R}{12}\big)^{\frac{1}{2}} \label{dotR}.
	\end{align}
\end{subequations}
Therefore, $H(t)\approx H(t_0)-\frac{t}{6} \alpha^{-1}$. Anisotropies of the CMB suggest $M\approx 10^{13}$ GeV \cite{NOJIRI201159, NOJIRI20171, defelice2010}, which corresponds to a $\alpha\approx 10^{11} M_P^{-2}$ with the reduced Planck mass $M_P=1/\sqrt{8\pi G}=2.44 \times 10^{18}$ GeV. Inflation starts with $\alpha R\gg 1$ and ends when the curvature decreases enough for the linear term to become relevant. 

We now briefly describe quantum aspects of the FLRW model of Starobinsky related to inflation. This model has been studied in the context of quantum cosmology by several authors. In \cite{hawkingluttrell,hawking-wu} qualitative properties of the no-boundary wave function were investigated. Further analysis using the tunneling from nothing proposal, including perturbations, can be found in \cite{vilenkin85,vilenkin88,mijic,Morris,Contreras}, where the probability distribution of the initial curvature for both boundary conditions were computed. These results were contrasted with those of the cubic curvature theory considered in \cite{Henk1994}. Approximate and exact solutions of the WDW equation for the pure quadratic curvature model were obtained in \cite{Kasper93,Pimentel94,Kenmoku96}. In \cite{Sanyal}, a preferred auxiliary variable to rewrite the quadratic curvature term was pointed out, with which the quantum theory allows the standard probability interpretation. More general setups have also been considered, e.g., quadratic curvature with non-standard couplings to a scalar field were studied in \cite{nahomi} and the quantization of general $f(R)$ theories in \cite{vazquez,universe7080288}. On the other hand, a perturbative analysis of the quadratic term was made in \cite{Mazzitelli} in connection with the issue of renormalization.
\subsubsection{Quantization}
In the Dirac quantization scheme, the condition that physical states are annihilated by the Hamiltonian operator $H_0$ constitutes the WDW equation. For the FLRW Starobinsky model (\ref{Hamiltonian}) it reads 
\begin{align}\label{starwdw}
	\frac{\kappa}{3} \left[\hbar^2 \left(\frac{\partial}{\partial a} \frac{\partial}{\partial \phi}-\frac{1}{2a} \frac{\partial}{\partial \phi}\right)+U(a,\phi)\right] \Psi(a,\phi)=0,
\end{align}
where we chose symmetric (Weyl) ordering for the non-commuting operators. The scalar potential is given by 
\begin{align}\label{potstar0}
	U(a,\phi)=\frac{9}{\kappa^2} \left( -k a \phi+\frac{1}{4 \alpha} a^2 (\phi-a)^2\right) 
\end{align}

We look for WKB solutions that hold in the classically inflationary region $\alpha R \gg 1$. Thus, following \cite{mijic} we consider the potential (\ref{potstar0}) to zeroth order in $\frac{a}{\phi}=\frac{1}{1+\frac{\alpha}{3} R}$, that is $\phi-a=\phi (1-\frac{a}{\phi})\approx \phi$. Substituting $\Psi=G(a,\phi) \exp [\frac{i}{\hbar} S(a,\phi)]$ into (\ref{starwdw}), we get, to zeroth order in Planck's constant,
\begin{align}\label{wkb0}
	(\partial_a S) \partial_\phi S-y^2 a^2 \phi^2=0,
\end{align}
where $y\equiv \frac{3}{2 \kappa \sqrt{\alpha}}$. This equation admits separable solutions of the form
\begin{align}
	S(a,\phi)=\pm \frac{2}{3} y \big[(a^3+c_1) (\phi^3+c_2)\big]^{\frac{1}{2}}
\end{align} 
Choosing $c_1=0=c_2$, the first order equation becomes 
\begin{align}
	a \partial_a G+\phi \partial_\phi G+G=0.
\end{align}
which is also separable. Thus, we obtain the WKB solutions
\begin{align}\label{wkb}
	\Psi_{0, \text{WKB}(1)}(a,\phi)=a^{-c-1} \phi^c \exp \left( \pm i \frac{2}{3 \hbar} y a^{\frac{3}{2}} \phi^{\frac{3}{2}}\right) , && \phi\gg a
\end{align}
The sub-index $0$ indicates that we neglected terms $\frac{a}{\phi}$ in the potential, whereas the label WKB$(1)$ indicates that we are considering only up to the first order term in the WKB series. The separation constant $c$ is determined by boundary conditions. 

Similarly, a WKB solution can be obtained for the region $a\gg \phi\ge 0$, that occurs when $R$ approaches $-\frac{3}{\alpha}$ from above. Considering (\ref{potstar0}) to zeroth order in $\frac{\phi}{a}$ we get the WKB solution $\Psi \propto  \exp \big(\pm \frac{3 i}{\hbar \kappa \sqrt{5 \alpha}} a^{\frac{5}{2}} \phi^{\frac{1}{2}}\big)$, which we will only use as a guide to introduce a first order correction $\frac{a}{\phi}$ to the solution (\ref{wkb}).

On the other hand, in the neighborhood of $\phi=a$, where the potential vanishes, we have $\partial_\phi (\partial_a-\frac{1}{2a}) \psi\approx 0$. The wave function there must be of the form $\Psi=a^{\frac{1}{2}} F(\phi)+G(a)$, with $F$ and $G$ arbitrary functions.

Therefore, a particular solution that connects both regions $\phi\gg a$ and $a\gg \phi$ is of the form
\begin{align}\label{wkball}
	\Psi_{0, \text{WKB}(1)}(a,\phi)=\frac{1}{\sqrt{a \phi}} \sin \left( \frac{1}{\hbar \kappa \sqrt{\alpha}} a^{\frac{3}{2}} \phi^{\frac{1}{2}} |\phi-a|\right) ,
\end{align}
and satisfies $\psi(0,\phi)=0$, $\psi(a,0)\propto a^2$. 

Oscillatory wave functions such as (\ref{wkb}) or (\ref{wkball}) predict a strong correlation between coordinates and momenta \cite{halliwell}. In this case we have, 
\begin{subequations}
\begin{align}
	p_a=-\frac{3}{\kappa} a \dot{\phi}=\frac{\partial S}{\partial a}=\pm \frac{3 \alpha^{-\frac{1}{2}}}{2 \kappa} a^{\frac{1}{2}} \phi^{\frac{3}{2}} \big(1-\frac{a}{\phi}\big), \label{pa} \\
	p_\phi=-\frac{3}{\kappa} a \dot{a}=\frac{\partial S}{\partial \phi}=\pm \frac{3 \alpha^{-\frac{1}{2}}}{2 \kappa}  a^{\frac{3}{2}} \phi^{\frac{1}{2}} \big(1-\frac{a}{\phi}\big) \label{pf},
\end{align}
\end{subequations}
respectively. The $\pm$ signs correspond to contracting/expanding phases. To zeroth order in $\frac{a}{\phi}$, (\ref{pf}) reproduces (\ref{RH}). On the other hand, multiplying (\ref{pa}) and (\ref{pf}) we get (\ref{eqf3}) ($k=0$), from which one derives (\ref{hinf}) and (\ref{dotR}). 

The analysis for $k=1$ is best performed using coordinates allowing separation of variables \cite{vilenkin85,mijic}. It is well known that through a conformal transformation and a field redefinition, (\ref{starlag}) can be expressed as the action for a minimally coupled scalar field on a curved spacetime \cite{cecotti,whitt}. In this FLRW setup the transformation reads $(a,\phi) \to (A,\varphi)=(\sqrt{a \phi},\frac{1}{c} \ln \frac{\phi}{a})$, with  $c^2=\frac{2\kappa}{3}$.
\begin{figure}[h!]
	\centering
	\includegraphics[scale=0.38]{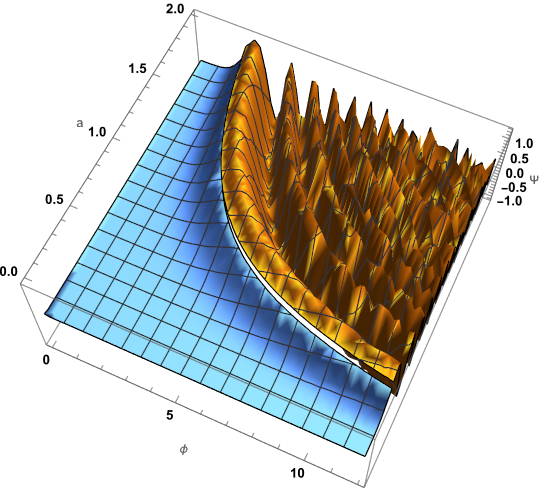}
	\caption{WKB solution (\ref{realsol}) for $k=1$, $c_1=0$.}
	\label{Fig-8}
\end{figure}
At zeroth order in $\frac{a}{\phi}$, there are two regions where the WKB approximation holds,
\begin{subequations}\label{wkbk1}
\begin{align}
	\Psi_{0, \text{WKB} (1)}&=\left( a \phi \Big(\frac{a \phi}{4 \alpha}-1\Big)\right)^{-\frac{1}{4}} \exp \Big(\pm i \frac{8 \alpha}{\kappa} \Big(\frac{a \phi}{4 \alpha} -1\Big)^\frac{3}{2}\Big), && a \phi\gg 4\alpha \\
	\Psi_{0, \text{WKB} (1)}&=\left( a \phi \Big(1-\frac{a \phi}{4 \alpha}\Big)\right)^{-\frac{1}{4}} \exp \Big(\pm \frac{8 \alpha}{\kappa} \Big(1-\frac{a \phi}{4 \alpha}\Big)^\frac{3}{2}\Big), && a \phi\ll 4\alpha.
\end{align}
\end{subequations}
By the standard matching conditions one gets (see Figure \ref{Fig-8})
\begin{subequations}\label{realsol}
	\begin{align}
		\Psi(a,\phi)&\approx \Big(\frac{a \phi}{2 \sqrt{\alpha}}\Big)^{-\frac{1}{2}} \left\lbrace c_1 \cos \Big[\frac{8 \alpha}{\hbar \kappa} \big(\frac{a \phi}{4\alpha}-1\big)^{\frac{3}{2}}+\frac{\pi}{4} \Big]
		+2 c_2 \sin \Big[\frac{8 \alpha}{\hbar \kappa} \big(\frac{a \phi}{4 \alpha}-1\big)^{\frac{3}{2}}+\frac{\pi}{4}\Big]\right\rbrace, \\
		\Psi(a,\phi)&\approx (a \phi)^{-\frac{1}{4}} \left\lbrace c_2 \exp \Big[\frac{-8 \alpha}{\hbar \kappa} \big(1-\frac{a \phi}{4 \alpha}\big)^{\frac{3}{2}}\Big]+c_1 \exp \Big[\frac{8 \alpha}{\hbar \kappa} \big(1-\frac{a \phi}{4 \alpha}\big)^{\frac{3}{2}}\Big]\right\rbrace.
	\end{align}
\end{subequations}
at $a \phi>4 \alpha$ and $a\phi<4 \alpha$, respectively.

\section{Supersymmetric actions}\label{sec3}
As mentioned in Section \ref{sec1}, we use a superfield approach. In general, superfields are functions on a spacetime enlarged by anti-commuting Grassmann coordinates, called superspace too (no relation to the superspace of quantum gravity). In superspace, supersymmetry transformations are nothing but special coordinate transformations. The formalism for one dimension has been detailed in \cite{garcia,ramirez2016,tesis} or \cite{ramirez} for arbitrary dimension. The obtained models possess local or time-dependent supersymmetry as well as time-reparametrization invariance.

The models described below provide a realization of either N=1 or N=2 1D supersymmetry algebra \cite{bellucci}. For N=1 SUSY there is only one real generator $Q$ such that $[Q,Q]_+=-2 P$, with $P$ the generator of time translations \cite{bellucci,ramirez2008}. With N=2 SUSY, we have two generators $[Q_1,Q_1]_+=-2 P=[Q_2,Q_2]$, $[Q_1,Q_2]_+=0$ or, in complex representation, $[S, \bar{S}]_+=-2P$, $S^2=0=\bar{S}^2$, where $S\equiv Q_1+i Q_2$, $\bar{S}\equiv Q_1-i Q_2$.

\subsection{N=1 SUSY}\label{n1}
N=1 superspace has local coordinates $\{t,\Theta\}$, where $t$ is an ordinary real number and $\Theta$ is a real Grassmann odd number: $\Theta=\Theta^*$ and $\Theta \Theta=0$. The model we proceed to study follows from the superfield Lagrangian \cite{tesis}
\begin{align}\label{streal}
	\mathcal{L}=\frac{3}{\kappa} \mathcal{NA}^3 \big(\mathcal{R}+\alpha \mathcal{R} \nabla_\Theta \mathcal{R}\big).
\end{align}
where $\mathcal{N}(t,\Theta)=N(t)-i \Theta \psi(t)$ is a scalar density \cite{wessbagger} whose components constitute the N=1 1D supergravity multiplet $\{N, \psi\}$. On the other hand,
\begin{subequations}
\begin{align}\label{A}
	\mathcal{A}(t,\Theta)&=a(t) (1+i\Theta \lambda (t)), \\
	\mathcal{R}(t,\Theta)&=-\frac{\dot{\lambda}}{N}-2\lambda \frac{\dot{a}}{aN}-\psi \frac{\dot{a}}{aN}+\Theta \big[\frac{1}{6} R_0+\frac{2i}{N}\lambda \dot{\lambda}-\frac{i}{N} \dot{\psi} \lambda-\frac{6i}{N}\psi \lambda \frac{\dot{a}}{a}-\frac{2i}{N}\psi \dot{\lambda}\big],
\end{align}
\end{subequations}
are the scale factor and the flat FLRW curvature superfields, respectively. $R_0$ is the flat scalar curvature (\ref{frwcurvature}). Note that $\mathcal{R}$ is nilpotent of degree two. In components, we get
\begin{align}\label{realsta}
	L_S\doteq \frac{3}{\kappa} a^3 \Big[\frac{1}{6} R+\frac{\alpha}{36} R^2-i \lambda \dot{\lambda}+\alpha \big(i \dot{\lambda} \ddot{\lambda}-i \frac{\ddot{a}}{a} \lambda \dot{\lambda}+7i \frac{{\dot{a}}^{2}}{a^2} \lambda \dot{\lambda}+2i \frac{\dot{a}}{a} \lambda \ddot{\lambda}\big)\Big] 
\end{align}
where $\doteq$ indicates the gauge $\mathcal{N}(t,\Theta)=1$.

For the Hamiltonian formulation we first rewrite the superfield action as follows
\begin{align}\label{staro1}
	\mathcal{L}=\frac{3}{\kappa} \mathcal{NA}^3 \big[\mathcal{R}+\alpha \Phi \nabla_{\Theta} \Phi+\alpha \Gamma (\Phi-\mathcal{R})\big],
\end{align}
where $\Phi=\eta+\Theta d$ and $\Gamma=g+i \Theta \gamma$ are odd and even superfields, respectively. In ({\ref{staro1}}), $\Gamma$ appears as a superfield multiplier enforcing the constraint $\Phi=\mathcal{R}$. However, since $\mathcal{R}$ depends on the supersymmetric covariant derivatives of $\mathcal{A}$ \cite{nephtali21}, $\Gamma$ is a dynamical superfield itself. In components, we get the Lagrangian
\begin{align}\label{n1lag}
	L=\frac{3}{\kappa} N a^3 \Big[\frac{1}{6} R_0+\alpha d^{2}+\alpha g (d-\frac{1}{6}R_0)-\frac{i}{N} \big(\dot{\psi} \lambda+\psi \lambda \frac{\dot{a}}{a}+\psi \dot{\lambda}+\lambda \dot{\lambda}\big)+\alpha \big(\frac{i}{N} \big(\eta \dot{\eta}+\dot{\psi} \lambda g\nonumber \\
	+\psi \lambda \frac{\dot{a}}{a} g +\psi \dot{\lambda} g+\lambda \dot{\lambda} g-2\lambda \gamma \frac{\dot{a}}{a}-\dot{\lambda} \gamma-\psi \gamma \frac{\dot{a}}{a}\big)-\eta \gamma i+3\lambda \eta d i+3\lambda \eta i g-\psi \eta i g\big)\Big]
\end{align}
Off-shell there are three bosons $\{a, g, d\}$ and three fermions $\{\lambda, \gamma, \eta\}$; $g$ and $\gamma$ are Lagrange multipliers leading to $d=R_0+...$ and $\eta=\dot{\lambda}+...$, respectively. We do not use the fermionic constraint for that would restore the higher-derivative kinetic term $\dot{\lambda} \ddot{\lambda}$, leading to third order equations of motion. Actually, (\ref{n1lag}) contains a kinetic term for $\gamma$ that can be made explicit by integrating by parts the term $\dot{\lambda} \gamma$. Therefore, we have three dynamical fermions $\{\lambda,\eta,\gamma\}$ satisfying first-order equations of motion (cf. the alternative formulation with two real fermions and second order equations in \cite{nephtali21}). On the other hand, one of the bosons is an auxiliary variable; we keep $g$ and eliminate $d$ by using its equation $d=-\frac{1}{2} (g+3 i \lambda \eta)$.

To make the Lagrangian look like (\ref{staraction}), we perform the following change of variables, 
\begin{align}\label{Qa}
	\phi=a(1-\alpha g), && \chi=-\alpha \gamma a+\lambda a (1-\alpha g)
\end{align}
Fully on-shell $\phi=a(1+\frac{\alpha}{3} R)$ plus terms containing fermions. A final re-scaling $\lambda\to a^{-1} \lambda$, $\chi\to a^{-1} \chi$ and $\eta \to a^{-\frac{3}{2}} \eta$ yields
\begin{align}\label{lagra1}
	L=\frac{3}{\kappa} \Big[-a \frac{\dot{\phi} \dot{a}}{N}-\frac{1}{4 \alpha} N a (\phi-a)^2+i \big(\alpha \eta \dot{\eta}+\dot{\phi} \psi \lambda a+\lambda \chi \frac{\dot{a}}{a}+\dot{\lambda} \chi+\psi \chi \dot{a}\big)
	+i N \big(\frac{3}{2}\lambda-\frac{\phi}{2a} \lambda-\frac{1}{a} \chi-\psi {a}^{\frac{1}{2}}+\phi \psi\big) \eta a^{\frac{1}{2}}\Big].
\end{align}
(\ref{lagra1}) depends on two physical bosons $\{a,\phi\}$ and three physical fermions $\{\lambda,\chi,\eta\}$. This is no surprise since in the original higher derivative action, following from (\ref{streal}), there are two bosonic degrees of freedom $\{a,\ddot{a}\}$ and three fermionic degrees of freedom $\{\lambda,\dot{\lambda},\ddot{\lambda}\}$ \cite{nephtali21}. A peculiarity of real physical fermions is that each one amounts to half a classical degree of freedom, since coordinate and momentum are given by the same quantity \cite{henneaux}. In the ordinary case, obtained by setting $\alpha=0$ in (\ref{streal}) and (\ref{staro1}), we only have one boson $a$ and one fermion $\lambda$. Therefore, by switching on the coupling constant $\alpha$, the number of bosons double while the number of fermions triple. In the complex case of Section \ref{arbi} below, the number of both bosons and fermions will triple.

The first-class Hamiltonian is $H=N (H_0+\psi S)+n p_N+\Sigma \pi_\psi$, where $p_N=\frac{\partial L}{\partial \dot{N}}\approx 0$, $\pi_\psi=\frac{\partial L}{\partial \dot{\psi}}\approx 0$. The Hamiltonian and supersymmetric constraints are, respectively,
\begin{align}
	H_0&=-\frac{\kappa}{3} \frac{p_\phi p_a}{a}+\frac{3}{4 \kappa \alpha} a (\phi-a)^2+\lambda \chi i \frac{p_\phi}{a^2}+\frac{3}{2\kappa}i \lambda \eta a^{-\frac{1}{2}} (\phi-3 a)-\frac{3}{\kappa} i \eta \chi {a}^{-\frac{1}{2}} \approx 0,\\
	S&=i \lambda p_a+i \chi a^{-1} p_\phi-\frac{3i}{\kappa} \eta {a}^{\frac{1}{2}} (\phi-a) \approx 0. \label{supercharge}
\end{align}
Since (\ref{lagra1}) is linear in fermionic velocities, we get second-class constraints
$\pi_\lambda-\frac{3i}{\kappa} \chi\approx 0$, $\pi_\chi \approx 0$, $\pi_\eta+\frac{3i}{\kappa} \alpha \eta \approx 0$, leading to the basic Dirac brackets
\begin{align}\label{n1brackets}
	\{a,p_a\}_D=1, && \{\phi,p_\phi\}_D=1, && \{\lambda,\chi\}_D=i \frac{\kappa}{3}, && \{\eta,\eta\}_D=-i \frac{\kappa}{6 \alpha}
\end{align}
Note that $\{\lambda,\lambda\}_D=0=\{\chi,\chi\}_D$.

The N=1 supersymmetry algebra is realized by the first-class constraints
\begin{align}
	\{S,S\}_D=-2 H_0, && \{S,H_0\}_D=0, && \{S,S\}_D=0.
\end{align}
Note that $S$ is purely imaginary.

\subsubsection{Quantization}\label{quantumn1}
Following the standard quantization rules, classical quantities are promoted to operators satisfying (anti)commutation relations determined by the rule $\{A,B\}_D=C \to \hat A \hat B\mp \hat B \hat A=i\hbar \hat C$ (the anticommutator is reserved for both $\hat A$ and $\hat B$ fermionic operators). Bosonic conjugate operators satisfy $[\hat q_A,\hat p_B]=i \hbar \hat{\delta}_{AB}$ and we use the standard position representation. In the following, the supergravity multiplet plays no role, thus, without the risk of confusion, from now on, $\Psi$ and $\psi_i$ will denote the quantum state of the universe and its components.

From (\ref{n1brackets}), the N=1 fermions satisfy the algebra
\begin{align}\label{n1algebra}
	[\lambda,\chi]_+=-\frac{\hbar \kappa}{3}, && \lambda^2=0=\chi^2, && \eta^2=\frac{\hbar \kappa}{12\alpha}
\end{align}
Defining $\gamma^0=(\frac{3}{\hbar \kappa})^{\frac{1}{2}} (\lambda+\chi)$, $ \gamma^1=(\frac{3}{\hbar \kappa})^{\frac{1}{2}} (\lambda-\chi)$ and $\gamma^2=(\frac{12 \alpha}{\hbar \kappa})^{\frac{1}{2}} \eta$ one can express (\ref{n1algebra}) as the more familiar 3D Clifford algebra of Lorentzian signature 
\begin{align}
	\gamma^\mu \gamma^\nu+\gamma^\nu \gamma^\mu=2\eta^{\mu \nu}, && (\mu=0,1,2)
\end{align}
whose irreducible representations are 2-dimensional, e.g., $\gamma^0=i \sigma_2$, $\gamma^1=\sigma_1$, $\gamma^2=\sigma_3$, using standard Pauli matrices $\sigma_i$. Solving for the fermions we get
\begin{align}\label{irrep}
	\lambda=\Big(\frac{\hbar \kappa}{3}\Big)^{\frac{1}{2}} \begin{pmatrix}
		0 & 1\\
		0 & 0
	\end{pmatrix}, && \chi=\Big(\frac{\hbar \kappa}{3}\Big)^{\frac{1}{2}}\begin{pmatrix}
		0 & 0\\
		-1 & 0
	\end{pmatrix}, && \eta=\Big(\frac{\hbar \kappa}{12 \alpha}\Big)^{\frac{1}{2}}\begin{pmatrix}
		1 & 0\\
		0 & -1
	\end{pmatrix}
\end{align}
which satisfy (\ref{n1algebra}). Considering that the classical fermions are defined real, one could have expected Hermitian matrices. For that feature one must resort to representations of higher dimension \mbox{\cite{henneaux}}.

The quantum state of the universe has two components
\begin{align}
	\Psi(a,\phi)=\begin{pmatrix}
		\psi_1(a,\phi) \\ i \psi_2(a,\phi)
	\end{pmatrix},
\end{align}
and is annihilated by the supersymmetric constraint operator
\begin{align}
	S=\lambda \hbar \partial_a+\chi a^{-1} \hbar \partial_\phi-\frac{3i}{\kappa} \eta {a}^{\frac{1}{2}} (\phi-a).
\end{align}

From the matrix equation $S\Psi=0$, we get the system of coupled equations
\begin{align}\label{n1eqs}
	\frac{\hbar}{a} \frac{\partial \psi_1}{\partial \phi}-y {a}^{\frac{1}{2}} (\phi-a) \psi_2=0, && 
	\hbar \frac{\partial \psi_2}{\partial a}+y {a}^{\frac{1}{2}} (\phi-a) \psi_1=0,
\end{align}
where $y=\frac{3}{2 \kappa \sqrt{\alpha}}$.

As in Section \ref{sec2}, we look for solutions valid in the inflationary domain $\alpha R\gg 1$. Let's consider equations (\ref{n1eqs}) retaining only the first term in the potential, that is $\phi(1-\frac{a}{\phi}) \approx \phi$. Substituting
\begin{align}\label{approx}
	\psi_1(a,\phi)=T_1 (a,\phi) \exp \big(\frac{1}{\hbar} S(a,\phi)\big), &&
	\psi_2(a,\phi)=T_2 (a,\phi) \exp \big(\frac{1}{\hbar} S(a,\phi)\big).
\end{align}
into (\ref{n1eqs}) we get, to zeroth order in $\hbar$,
\begin{align}\label{first}
	T_1 \partial_\phi S=y a^{\frac{3}{2}} \phi T_2 && T_2 \partial_a S=-y a^{\frac{1}{2}} \phi T_1
\end{align}
Solving for $\frac{T_1}{T_2}$ from the two equations and equating we get
\begin{align}\label{seq}
	(\partial_a S) \partial_\phi S+y^2 a^2 \phi^2=0,
\end{align}
which can be solved by separation of variables. A solution is
\begin{align}
	S(a,\phi)=\pm i \frac{2}{3} y a^{\frac{3}{2}} \phi^{\frac{3}{2}}.
\end{align}
Consequently, from (\ref{first})
\begin{align}\label{ts}
	T_1=\mp i \phi^{\frac{1}{2}} T_2
\end{align}
Next, to first order in $\hbar$ we get
\begin{align}
	\partial_\phi T_1=0, && \partial_a T_2=0.
\end{align}
which combined with (\ref{ts}) determine the $T$'s. Thus we get
\begin{subequations}\label{soln1}
	\begin{align}
		(\psi_1)_0(a,\phi)&=c_1 \exp \Big(\pm i \frac{\alpha^{-\frac{1}{2}}}{\hbar \kappa} a^{\frac{3}{2}} \phi^{\frac{3}{2}}\mp i \frac{\pi}{2}\Big), \\
		(\psi_2)_0 (a,\phi)&=c_1 \phi^{-\frac{1}{2}} \exp \Big(\pm i \frac{\alpha^{-\frac{1}{2}}}{\hbar \kappa \sqrt{\alpha}} a^{\frac{3}{2}} \phi^{\frac{3}{2}}\Big),
	\end{align}
\end{subequations}
where $c_1$ is a constant. The sub-index $0$ remind us that we discarded terms $\frac{a}{\phi}$ in the potential. 
\begin{figure}[h!]
	\centering
	\subfigure[]{\includegraphics[scale=0.5]{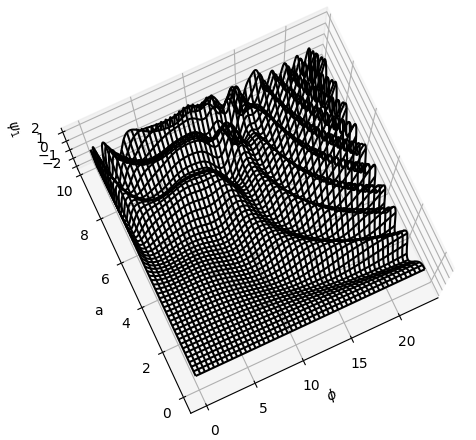}}
	\subfigure[]{\includegraphics[scale=0.5]{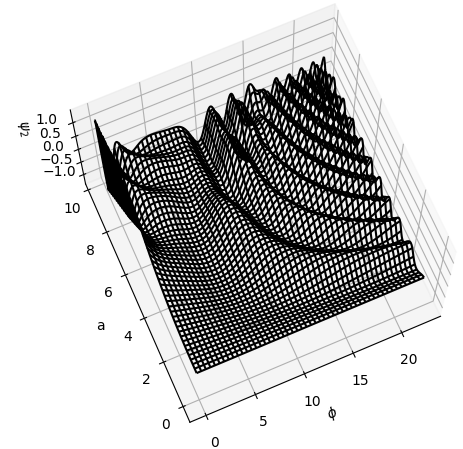}}
	\caption{Numerical solution to equations (\ref{n1eqs}): (a) $\psi_1(\phi,a)$, (b) $\psi_2(\phi,a)$ for $\alpha=10^3$ and $\kappa=1$.}
	\label{n1sol}
\end{figure}
Although solutions (\ref{soln1}) have been obtained in WKB fashion, they solve exactly the supersymmetric equations (\ref{n1eqs}) in the referred approximation. That is why we do not write an extra WKB label as in solutions (\ref{wkb}).

An analogous procedure for the other regime $a\gg \phi\ge 0$ yields
\begin{subequations}\label{wkbaggf}
	\begin{align}
		\psi_1(a,\phi)&=\sqrt{5} c_2 \exp \Big(\pm \frac{3 i}{\sqrt{5 \alpha} \hbar \kappa} a^{\frac{5}{2}} \phi^{\frac{1}{2}}\pm i \frac{\pi}{2}\Big), \\
		\psi_2(a,\phi)&=c_2 \phi^{-\frac{1}{2}} \exp \Big(\pm \frac{3 i}{\sqrt{5 \alpha} \hbar \kappa} a^{\frac{5}{2}} \phi^{\frac{1}{2}}\Big).
	\end{align}
\end{subequations}
These solutions indicate us how powers of $a$ and $\phi$ in the phase of the solutions change as we move from one regime to the other. 

Considering that, except for the origin, both (\ref{soln1}) and (\ref{wkbaggf}) are finite at $\phi=a$, from (\ref{n1eqs}) we get the conditions
\begin{align}
	\frac{\partial \psi_1}{\partial \phi}=0, && 
	\frac{\partial \psi_2}{\partial a}=0,
\end{align}
along the curve $\phi=a$. 

Thus, introducing a first order perturbation, our asymptotic solutions are
\begin{subequations}\label{approxn1}
	\begin{align}
		\psi_1(a,\phi)&=c \cos \Big(\frac{1}{\sqrt{\alpha} \hbar \kappa} a^{\frac{3}{2}} \phi^{\frac{1}{2}} |\phi-a|\Big), \\
		\psi_2(a,\phi)&=\mp c \phi^{-\frac{1}{2}} \sin \Big(\frac{1}{\sqrt{\alpha} \hbar \kappa} a^{\frac{3}{2}} \phi^{\frac{1}{2}} |\phi-a|\Big),
	\end{align}
\end{subequations}
We can use the asymptotic solutions to deduce boundary conditions for a numerical solution of (\ref{n1eqs}): $\psi_1(a,0)=\sqrt{5}$ and $\psi(0,\phi)=0$. A numerical solution of the coupled equations (\ref{n1eqs}), shown in Figure \ref{n1sol}, exhibits the expected oscillatory behavior (\ref{approxn1}).

We must stress that the wave functions (\ref{approxn1}) are fixed up to a normalization constant, whereas those obtained for the non-supersymmetric model allow for different powers in the quantum prefactor.

\section{N=2 SUSY}\label{quadratic}
N=2 superspace has local coordinates $\{t, \Theta, \bar{\Theta}\}$ and scalar density $\mathcal{N}=N (1+i \Theta \bar{\psi}+i \bar{\Theta} \psi)$, where an over bar denotes complex conjugation. The N=2 scale factor and curvature superfields are given by
\begin{subequations}
	\begin{align}
		\mathcal{A}&=a \big(1+i\Theta \bar{\lambda}+i\bar{\Theta} \lambda-\Theta \bar{\Theta} (s-\sqrt{k} a^{-1}-\lambda \bar{\lambda})\big), \label{scale} \\
		\mathcal{R}&=s+\Theta \Big(\frac{\dot{\bar{\lambda}}}{N}+2\bar{\lambda} \frac{\dot{a}}{a N}-\bar{\psi} \frac{\dot{a}}{a N}-\psi \bar{\psi} \bar{\lambda} i+\bar{\psi} \frac{\sqrt{k}}{a} i-\bar{\psi} i s-2\bar{\lambda} i s+\bar{\lambda} \frac{\sqrt{k}}{a} i\Big) \nonumber \\
		&\qquad -\bar{\Theta} \Big(\frac{\dot{\lambda}}{N}+2\lambda \frac{\dot{a}}{a N}-\psi \frac{\dot{a}}{a N}+\psi \bar{\psi} \lambda i-\psi \frac{\sqrt{k}}{a} i+\psi i s+2\lambda i s-\lambda \frac{\sqrt{k}}{a} i\Big) \nonumber \\
		&\qquad +\Theta \bar{\Theta} \Big(\frac{1}{6} R+2s^2-3\frac{\sqrt{k}}{a} s+2\frac{\sqrt{k}}{a} \psi \bar{\psi}-\frac{i}{N} (\dot{\psi} \bar{\lambda}+\dot{\bar{\psi}} \lambda)-\frac{2 i}{N} (\psi \dot{\bar{\lambda}}+\bar{\psi} \dot{\lambda}) \nonumber \\
		&\qquad -\frac{6i\dot{a}}{a N} (\psi \bar{\lambda}+\bar{\psi} \lambda)-\frac{2i}{N} (\lambda \dot{\bar{\lambda}}+\bar{\lambda} \dot{\lambda})+2 \frac{\sqrt{k}}{a} (\psi \bar{\lambda}-\bar{\psi} \lambda)-2s (\psi \bar{\lambda}-\bar{\psi} \lambda) \nonumber \\
		&\qquad +5\lambda \bar{\lambda} \frac{\sqrt{k}}{a}-2\psi \bar{\psi} s-8s \lambda \bar{\lambda}-4\psi \bar{\psi} \lambda \bar{\lambda}\Big). \label{supercurvature}
	\end{align}
\end{subequations}
The special form of (\ref{scale}) was chosen to simplify the lowest component of (\ref{supercurvature}). We will also make use of the following generic superfields 
\begin{subequations}\label{generic}
\begin{align}
\Phi&=f+i\Theta \bar{\eta}+i \bar{\Theta} \eta+\Theta \bar{\Theta} G,\\
\Gamma&=g+i \Theta \bar{\gamma}+i \bar{\Theta} \gamma+\Theta \bar{\Theta} M.
\end{align}
\end{subequations}

\subsection{F($\mathcal{R}$)}\label{secfdr}
The N=2 curvature superfield (\ref{supercurvature}) has even parity and we can define nontrivial $F(\mathcal{R})$ actions, analogous to $F(R)$ theories of gravity \cite{defelice2010}. An study of this action was done in \cite{tesis}. Here we give a summary of results, using a convenient set of coordinates, and also identify the superpotential of the theory. See e.g., \cite{Ketov13,Diamandis17} for the embedding of $F(R)$ theories into 4D supergravity.

The $F(\mathcal{R})$ action can be written with one extra superfield as follows \cite{tesis}
\begin{align}\label{superfdrphi}
	\mathcal{L}_F=\frac{3}{\kappa} \mathcal{NA}^3 \big(\mathcal{R} F'(\Phi)-\Phi F'(\Phi)+F(\Phi)\big)
\end{align}
with functions of superfields defined by Taylor series, $F(\Phi)=F(f)+F'(f) (\Phi-f)+\frac{1}{2} F''(f) (\Phi-f)^2$. In components, we get, after eliminating auxiliary fields,
\begin{align}\label{fdrlag}
	L\doteq \frac{3}{\kappa} a^3 \Big[\frac{1}{6} F' R+2s^2F'-3s F-3 \frac{\sqrt{k}}{a} (s F'-F)-3F \lambda \bar{\lambda}+F'' \big(i (\eta \dot{\bar{\lambda}}+\bar{\eta} \dot{\lambda})-2i \frac{\dot{a}}{a} (\lambda \bar{\eta}+\bar{\lambda} \eta) \nonumber \\
	-s (\lambda \bar{\eta}-\bar{\lambda} \eta)-\frac{\sqrt{k}}{a} (\lambda \bar{\eta}-\bar{\lambda} \eta)+3s (\lambda \bar{\eta}-\bar{\lambda} \eta)\big)+F' \big(i (\lambda \dot{\bar{\lambda}}+\bar{\lambda} \dot{\lambda})-\frac{\sqrt{k}}{a} \lambda \bar{\lambda}+4 \lambda \bar{\lambda} s+\eta \bar{\eta}\big)\Big].
\end{align}
On-shell $\eta=\dot{\lambda}+...$, which leads to the quadratic kinetic term $\dot{\lambda} \bar{\eta}=\dot{\lambda} \dot{\bar{\lambda}}$, leading to second-order equations of motion. Thus, there are two fermionic degrees of freedom. In the following section, $\lambda$ will satisfy a third order equation of motion.

The Hamiltonian formulation takes a simpler form in the following variables
\begin{align}\label{newcor}
	\phi \equiv aF'(s), && \chi\equiv a \eta F''(s)+\lambda a F'(s).
\end{align}
After a further re-scaling we are ready to compute the Legendre transformation. Bosonic momenta are defined in the usual way; for fermions, however, we use the notation $\pi_\lambda \equiv \partial L/\partial \dot{\bar{\lambda}}$, $\pi_{\bar{\lambda}} \equiv -\partial L/\partial \dot{\lambda}$, and so on. The momenta conjugate to the 1D supergravity multiplet vanish as constraints $p_N\approx0$, $\pi_\psi\approx0\approx \pi_{\bar{\psi}}$.
Performing the Legendre transformation $H=\dot{a} p_a+\dot{\phi} p_\phi-\dot{\lambda} \pi_{\bar{\lambda}}+\dot{\bar{\lambda}} \pi_\lambda-\dot{\chi} \pi_{\bar{\chi}}+\dot{\bar{\chi}} \pi_\chi-L$ we get a first-class Hamiltonian of the form $H=N (H_0+\frac{1}{2} \psi \bar{S}-\frac{1}{2} \bar{\psi} S)+n p_N+\Sigma \pi_\psi-\bar{\Sigma} \pi_{\bar{\psi}}$, with the Hamiltonian and supersymmetric constraints given by
\begin{align}
	H_0&=-\frac{\kappa}{3} a^{-1} p_\phi p_a+V(a,\phi)-i (\lambda \bar{\chi}+\bar{\lambda} \chi) {a}^{-2}p_\phi+\frac{3}{\kappa} \big(-a^{-1} F''^{-1} \chi \bar{\chi} \nonumber \\
	&+(\lambda \bar{\chi}-\bar{\lambda} \chi) \big(\frac{\sqrt{k}}{a}-2 \phi+F''^{-1} F'\big)+\lambda \bar{\lambda} (3aF-\sqrt{k}F'-a F''^{-1} F'^2) \big), \label{fdrhamil}\\
	-S&=\lambda (i p_a-W_a)+\chi a^{-1} (i p_\phi-W_\phi)-\frac{3}{\kappa} a^{-1} \lambda \bar{\lambda} \chi, \\
	\bar{S}&=\bar{\lambda} (i p_a+W_a)+\bar{\chi} a^{-1} (i p_\phi+W_\phi)+\frac{3}{\kappa} a^{-1} \lambda \bar{\lambda} \bar{\chi}.
\end{align}
In the above expressions, $F$ and its derivatives $F', F''$ must be evaluated at $s(\phi,a)$. The scalar potential in (\ref{fdrhamil}) is
\begin{align}
	V(a,s)=\frac{3}{\kappa} \big[-k a F'+3 \sqrt{k} a^2 (s F'-F)+a^3 (3 s F-2 s^2 F')\big]_{s=s(a,\phi)}. \label{scalarpot}
\end{align}
It can be expressed in terms of a more fundamental quantity, the superpotential
\begin{align}
		W(a,s)&=\frac{3}{\kappa} \big[\sqrt{k} a^2 F'-a^3 (s F'-F)\big], \label{superfdr}
\end{align}
according to $V=\frac{\kappa}{3} (-a^{-2} F''^{-1} W_a W_s+a^{-3} F' F''^{-2} W_s^2)$. Sub-indices indicate partial derivatives with respect to the mini-superspace coordinates $q_A=\{a,s\}$. The metric $G^{AB}$ can be read off the Hamiltonian by expressing the kinetic bosonic term as $\frac{1}{2} G^{AB} p_A p_B$. In terms of the variables $\{a,\phi\}$ we have $V(a,\phi)=-\frac{\kappa}{3} a^{-1} W_a W_\phi$, with $W_a=\frac{3}{\kappa} (\sqrt{k} \phi-2 \phi a s+3 a^2 F)|_{s(a,\phi)}$, $W_\phi=\frac{3}{\kappa} (\sqrt{k} a-a^2 s)|_{s(a,\phi)}$.

The second class constraints, $\pi_\lambda \approx 0$, $\pi_{\bar{\lambda}}\approx 0$, $\pi_\chi+\frac{3i}{\kappa} \lambda\approx 0$, $\pi_{\bar{\chi}}-\frac{3i}{\kappa} \bar{\lambda}\approx 0$ lead to the non-vanishing Dirac brackets
\begin{align}\label{fdrbrackets}
\{a,p_a\}_D=1, && \{\phi,p_\phi\}_D=1, && \{\lambda,\bar{\chi}\}_D=-i \frac{\kappa}{3}, && \{\bar{\lambda},\chi\}_D=-i \frac{\kappa}{3}.	
\end{align}

\subsubsection{Quantization}\label{quantumfdr}
From (\ref{fdrbrackets}) and the Correspondence Principle, the complex fermions satisfy the anti-commutators
\begin{align}\label{fdralgebra}
	[\lambda,\bar{\chi}]_+=\frac{\hbar \kappa}{3}=[\bar{\lambda},\chi]_+.
\end{align}
The fermionic sector can be represented with a finite dimensional space of states \cite{ramirez2016}. First, we define the following creation/annihilation operators 
\begin{subequations}\label{fdranticon}
\begin{align}
A&=\Big(\frac{3}{2\hbar \kappa}\Big)^{\frac{1}{2}} (\lambda-\chi), && A^\dagger=\Big(\frac{3}{2\hbar  \kappa}\Big)^{\frac{1}{2}} (\bar{\lambda}-\bar{\chi}), \label{aop}\\
B&=\Big(\frac{3}{2\hbar \kappa}\Big)^{\frac{1}{2}} (\lambda+\chi), && B^\dagger=\Big(\frac{3}{2\hbar  \kappa}\Big)^{\frac{1}{2}} (\bar{\lambda}+\bar{\chi}), \label{bop}
\end{align}	
\end{subequations}
in terms of which (\ref{fdralgebra}) is expressed as the algebra of two fermionic oscillators, one with the wrong sign,
\begin{align}
	[A,A^\dagger]_+=-1, && [B,B^\dagger]_+=1.
\end{align}

A subsidiary vacuum state $\ket{1}$ is defined such that $A\ket{1}=0=B\ket{1}$. Acting on it with the creation operators, we obtain the following independent states: $\ket{1}$, $\ket{2}=A^\dagger \ket{1}$, $\ket{3}=B^\dagger \ket{2}$, $\ket{4}=A^\dagger B^\dagger \ket{3}$. The vectors $\ket{i=1,2,3,4}$ are orthogonal and their norms are related by $\braket{1|1}=-\braket{2|2}=\braket{3|3}=-\braket{4|4}$. 

The space of states induces a matrix representation. Since there are negative norm states, the identity operator is given by $1=\sum_i \braket{i|i} \ket{i} \bra{i}$. Let $X$ be an arbitrary operator, we define its matrix representation such that $X=\sum_{i j} \braket{i|i} \braket{j|j} \ket{i} \braket{i|X|j} \bra{j}\equiv \sum_{i j} X_{ij} \braket{i|i} \ket{i} \bra{j}$. This convention respects matrix multiplication, $(X Y)_{il}=\sum_j X_{ij} Y_{jl}$, which is convenient for computations. Note that the  Hermitian conjugate operator matrix is now given by $(X^\dagger)_{ij}=X_{ji}^* \braket{i|i} \braket{j|j}$. The induced matrix representation is
\begin{subequations}\label{fdrladder}
\begin{align}
	A&=\sigma_+ \otimes 1, && A^\dagger=-\sigma_- \otimes 1, \\
	B&=\sigma_3 \otimes \sigma_+, && B^\dagger=\sigma_3 \otimes \sigma_-.	
\end{align}
\end{subequations}
where $\sigma_\pm=\frac{1}{2} (\sigma_1\pm i \sigma_2)$.

An arbitrary state has four components,
\begin{align}
	\Psi(a,\phi)=\begin{pmatrix}
		\psi_1(a,\phi) \\ \psi_2(a,\phi) \\ \psi_3(a,\phi) \\ \psi_4(a,\phi).
	\end{pmatrix}
\end{align}

Choosing anti-symmetric ordering for fermions, the supercharge operators read
\begin{align}
	S&=\lambda (-\hbar \partial_a+W_a)+\chi {a}^{-1} (-\hbar \partial_\phi+W_\phi)+\frac{1}{6\kappa} {a}^{-1} \lambda [\bar{\lambda}, \chi], \\
	\bar{S}&=\bar{\lambda} (\hbar \partial_a+W_a)+\bar{\chi} {a}^{-1} (\hbar \partial_\phi+W_\phi)-\frac{1}{6 \kappa} {a}^{-1} \bar{\lambda} [\lambda, \bar{\chi}].
\end{align}

The quantum supersymmetric constraints $S\Psi=0$, $\bar S\Psi=0$ yield six first order PDE's for the components. The wave functions associated to the empty $\psi_1$ and the filled states $\psi_4$ are determined up to normalization,
\begin{align}\label{N1analytic}
	\psi_1(a,\phi)&=c_1 a^{\frac{1}{2}} \exp \big[-\frac{3}{\hbar \kappa} \big(\sqrt{k} a \phi-a^3 \big(s F'(s)-F(s)\big)\big)|_{s(a,\phi)}\big], \\
	\psi_4(a,\phi)&=c_4 a^{\frac{1}{2}} \exp \big[\frac{3}{\hbar \kappa} \big(\sqrt{k} a \phi-a^3 \big(s F'(s)-F(s)\big)\big)|_{s(a,\phi)}\big]
\end{align}

The remaining components satisfy a system of first order coupled equations. They can be written more compactly in terms of $\psi_\pm \equiv \psi_3\pm \psi_2$,
\begin{align}\label{fdrcoupled}
	\frac{\hbar}{a} \partial_\phi  \psi_- -W_{a} \psi_+=0, && \hbar \big(\partial_a+\frac{1}{2a}\big)\psi_+ -\frac{1}{a} W_{\phi} \psi_-=0.
\end{align} 

\begin{figure}
	\centering
	\includegraphics[scale=0.28]{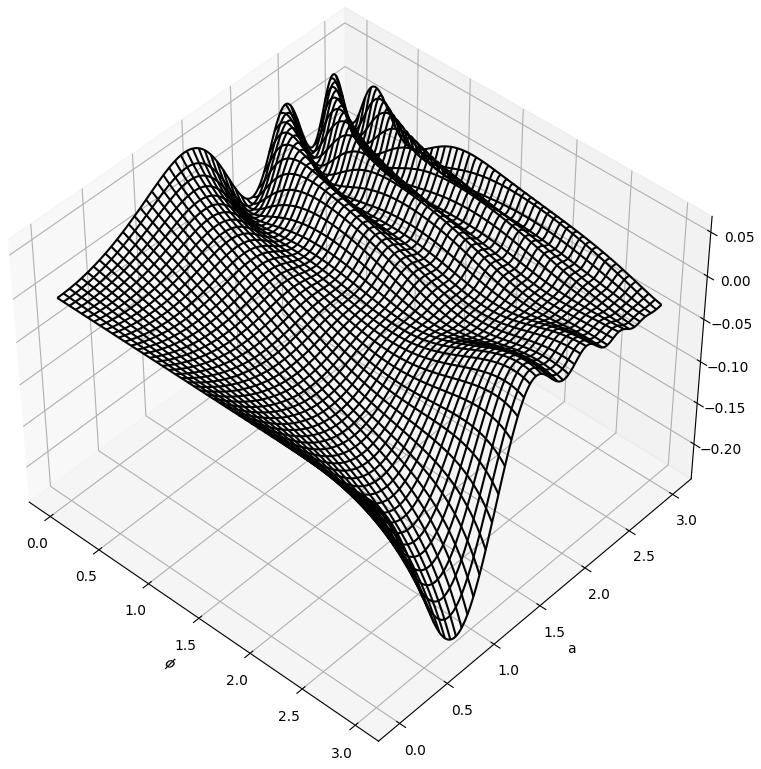}
	\caption{Numerical solution to (\ref{fdrcoupled}) for $\psi_+$. It grows exponentially in the region of unbounded negative potential.}
	\label{quadraticfdr}	
\end{figure}

As an example, we take $F(\mathcal{R})=\mathcal{R}+\alpha^{\frac{1}{2}} \mathcal{R}^2$. It turns out that the bosonic part of the corresponding Lagrangian looks like the Starobinsky model but only for small curvature. Moreover, the value of $R$ is bounded from above \cite{tesis}. The superpotential is
\begin{align}
	W=-\frac{3}{\kappa} \sqrt{\alpha} a^3 s^2=-\frac{3}{4 \kappa \sqrt{\alpha}} a (\phi-a)^2
\end{align}
and (\ref{fdrcoupled}) become
\begin{subequations}
\begin{align}
	\frac{\hbar}{a} \partial_\phi  \psi_- +\frac{3}{2 \kappa \sqrt{\alpha}} \big[\frac{1}{2} (\phi-a)^2-a (\phi-a)\big] \psi_+=0, \\
	\hbar \big(\partial_a+\frac{1}{2a}\big)\psi_+ +\frac{3}{2 \kappa \sqrt{\alpha}} (\phi-a) \psi_-=0,
\end{align}
\end{subequations}
which have the form of the N=1 equations (\ref{n1eqs}) (if we identify $\psi_-\to \psi_1$ and $\psi_+ \to \psi_2$) except for the quadratic term in the first equation. A numerical solution to $\psi_+$ is shown in Figure \ref{quadraticfdr}.

On the other hand, the exact solutions are,
\begin{align}
	\psi_\pm=a^{\frac{1}{2}} \exp \big[\pm \frac{3}{\hbar \kappa} \big(\sqrt{k} a \phi-\frac{1}{4\sqrt{\alpha}} a (\phi-a)^2-\Lambda a^3 \big)\big]
\end{align}
where we included a constant $\Lambda$ in the superpotential that makes square integrable one of the analytic solutions.

\subsection{$(\nabla \mathcal{R})^2-F(\mathcal{R})$}\label{arbi}
\begin{figure}[h!]
	\centering
	\includegraphics[scale=0.5]{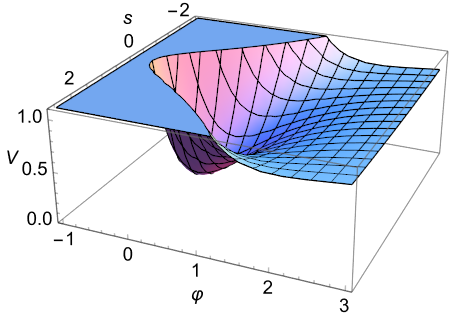}
	\caption{Multi-field potential of the N=2 model $V(\varphi,s)\propto (1-e^{-2 c\varphi})^2+e^{-2c \varphi} s^2$.}
	\label{Fig-5}
\end{figure}

It turns out that, by considering superspace covariant derivatives of $\mathcal{R}$, more specifically a superfield kinetic term, we obtain an action leading to third order equations of motions for $\lambda$ as with the N=1 model of Section \ref{n1}. Furthermore, the bosonic part contains the quadratic curvature term, not bounded from above as with $\mathcal{R}^2$ above. The superfield action is given by \cite{nephtali21}
\begin{align}\label{starsuper}
	\mathcal{L}=\frac{3}{\kappa} \mathcal{NA}^3 \Big(\mathcal{R}+\alpha \nabla_{\bar{\Theta}} \mathcal{R} \nabla_\Theta \mathcal{R}-\frac{4}{3} \alpha \mathcal{R}^3\Big).
\end{align}
where the last cubic term was the simplest choice making stable the potential along the $s$-direction. Indeed, the k=0 Lagrangian is (in the gauge $\mathcal{N}(t,\Theta,\bar{\Theta})=1$)
\begin{align}\label{lagrangianboson}
	L&\doteq \frac{3}{\kappa} a^3 \Big[\frac{1}{6} R+\frac{\alpha}{36} R^2+\alpha \dot{s}^{2}-{s}^{2}+s \lambda \bar{\lambda} +i (\lambda \dot{\bar{\lambda}}+\bar{\lambda} \dot{\lambda})+\alpha \big(i (\ddot{\bar{\lambda}} \dot{\lambda}+\ddot{\lambda} \dot{\bar{\lambda}}) \nonumber\\
	&\qquad +4 \frac{\dot{a}}{a} \dot{s} \lambda \bar{\lambda}+s \dot{\lambda} \dot{\bar{\lambda}}+\dot{\lambda} \dot{\bar{\lambda}} \lambda \bar{\lambda}+2 s (\lambda \ddot{\bar{\lambda}}-\bar{\lambda} \ddot{\lambda})+i \frac{\ddot{a}}{a} (\lambda \dot{\bar{\lambda}}+\bar{\lambda} \dot{\lambda})-2i \frac{\dot{a}}{a} (\lambda \ddot{\bar{\lambda}}+\bar{\lambda} \ddot{\lambda}) \nonumber\\
	&\qquad -7i \frac{\dot{a}^2}{a^2} (\lambda \dot{\bar{\lambda}}+\bar{\lambda} \dot{\lambda})+10 s \frac{\dot{a}}{a} (\lambda \dot{\bar{\lambda}}-\bar{\lambda} \dot{\lambda})+4 \frac{\ddot{a}}{a} s \lambda \bar{\lambda}+8 \frac{\dot{a}^{2}}{a^2} s \lambda \bar{\lambda}+\dot{s} (\lambda \dot{\bar{\lambda}}-\bar{\lambda} \dot{\lambda})\big)\Big], 
\end{align}
Thus, we get quadratic curvature and a massive scalar field. In the Einstein frame this model possesses the multi-field potential shown in Figure \ref{Fig-5}. We are mainly interested in the Starobinsky part and will consider $s$ as an spectator (c.f., multi field inflationary potential in supergravity \cite{Ketov_2022}).

As in the previous sections, we first rewrite (\ref{starsuper}) as an ordinary second-order theory using two additional even superfields $\Gamma$ and $\Phi$ (\ref{generic}), as follows,
\begin{align}\label{l2tres}
	\mathcal{L}=\frac{3}{\kappa} \mathcal{NA}^3 \big(\mathcal{R}+\alpha \nabla_{\bar{\Theta}} \Phi \nabla_\Theta \Phi-\alpha F(\Phi)+\alpha \Gamma (\Phi-\mathcal{R})\big),
\end{align}
where we allowed for an arbitrary function $F$. To obtain (\ref{starsuper}) we set $F(\Phi)=\frac{4}{3} \Phi^3$.

For simplicity, we use the notation $\mathcal{R}=s+\Theta \rho-\bar{\Theta} \bar{\rho}+\Theta \bar{\Theta} T$ for the components of (\ref{supercurvature}) in the next discussion. The part of the component Lagrangian following from (\ref{l2tres}), proportional to $\alpha$, is
\begin{align}
	L_\alpha=\frac{3}{\kappa} Na^3 \Big[\frac{\dot{f}^2}{N^2}+(G-T) g+G^2-F' G+M (f-s)+3 (\frac{\sqrt{k}}{a}-s) (f g-g s-F)+F'' \eta \bar{\eta} \nonumber \\
	-i \frac{1}{N} (\eta \dot{\bar{\eta}}+\bar{\eta} \dot{\eta})-2i \frac{\dot{f}}{N} (\psi \bar{\eta}+\bar{\psi} \eta)-2\psi \bar{\psi} \eta \bar{\eta}-\eta \bar{\gamma}+\bar{\eta} \gamma+\bar{\rho} \bar{\gamma} i+\rho \gamma i+3i \frac{\dot{f}}{N} (\lambda \bar{\eta}+\bar{\lambda} \eta) \nonumber \\
	+3 (F-f g+g s) \lambda \bar{\lambda}-3g (\lambda \bar{\eta}-\bar{\lambda} \eta)-3 g i (\lambda \rho+\bar{\lambda} \bar{\rho})+3f (\bar{\lambda} \gamma-\lambda \bar{\gamma})+3s (\lambda \bar{\gamma}-\bar{\lambda} \gamma) \nonumber \\
	+f (\bar{\psi} \gamma-\psi \bar{\gamma})+(\psi \bar{\gamma}-\bar{\psi} \gamma) s+F' (\psi \bar{\eta}-\bar{\psi} \eta)-g (\psi \bar{\eta} -\bar{\psi} \eta)-i g (\psi \rho+\bar{\psi} \bar{\rho})+3\frac{\sqrt{k}}{a} \eta \bar{\eta} \nonumber \\
	+3G (\bar{\lambda} \eta-\lambda \bar{\eta})-3\lambda \bar{\lambda} \eta \bar{\eta}-3\eta \bar{\eta} s+3F' (\lambda \bar{\eta}-\bar{\lambda} \eta)+3 (\psi \bar{\lambda}-\bar{\psi} \lambda) (F+g s-g f)\big]
\end{align}
Off-shell we have six bosons $a, d, g, s, M, G$ and three complex fermions $\lambda, \eta, \gamma$. At the superfield level $\Gamma$  enforces the constraint $\Phi=\mathcal{R}$ which is equivalent to the simultaneous equations: $f=s$, $i \bar{\eta}=\rho$, $i \eta=-\bar{\rho}$, $G=T$. We shall use only a subset of the equations of motion. First, $M$ is a multiplier enforcing the constraint $f-s=0$, thus we send $d\to s$. Next, we use the equation for $G$,  $2G=F'-g+3 (\lambda \bar{\eta}-\bar{\lambda} \eta)$, to eliminate it. In this way, we stay with an action depending on three bosons $\{a, s, g\}$ and three complex fermions $\{\lambda,\gamma,\eta\}$. Thus, the fermionic degrees of freedom triple just as with the N=1 model.

Substituting the components of $\mathcal{R}$ (\ref{supercurvature}), we get the full component Lagrangian
\begin{align}\label{fulllag}
	L=\frac{3}{\kappa} N a^3 \Big[\frac{1}{6} R-{s}^{2}-\frac{i}{N} (\dot{\psi} \bar{\lambda}+\dot{\bar{\psi}} \lambda)-i \frac{\dot{a}}{aN} (\psi \bar{\lambda}+\bar{\psi} \lambda) -\frac{i}{N} (\psi \dot{\bar{\lambda}}+\bar{\psi} \dot{\lambda})+\frac{i}{N} (\lambda \dot{\bar{\lambda}}+\bar{\lambda} \dot{\lambda}) \nonumber\\
	-2\frac{\sqrt{k}}{a} (\psi \bar{\lambda}-\bar{\psi} \lambda)+(s-\frac{\sqrt{k}}{a}) \lambda \bar{\lambda}+2\psi \bar{\psi} \lambda \bar{\lambda}+\alpha \Big(\frac{\dot{s}^{2}}{N^2}-\frac{1}{6} R g-\frac{1}{4} (F'-g)^2-2{s}^{2} g \nonumber\\
	+3 \frac{\sqrt{k}}{a} (g s-F)+3 sF-\frac{i}{N} (\eta \dot{\bar{\eta}}+\bar{\eta} \dot{\eta})-2 \frac{i \dot{s}}{N} (\psi \bar{\eta}+\bar{\psi} \eta)-2\psi \bar{\psi} \eta \bar{\eta}+\frac{3}{2} F' (\lambda \bar{\eta}-\bar{\lambda} \eta)
	\nonumber \\
	-4\lambda \bar{\lambda} s g+\frac{3}{2}\lambda \bar{\lambda} \eta \bar{\eta}+F'' \eta \bar{\eta}+\frac{i g}{N} (\dot{\psi} \bar{\lambda} +\dot{\bar{\psi}} \lambda)+\frac{i}{N} g (\psi \dot{\bar{\lambda}}+\bar{\psi} \dot{\lambda})+\frac{\dot{a}}{aN} g i (\psi \bar{\lambda}+\bar{\psi} \lambda) \nonumber\\
	+3\eta \bar{\eta} \frac{\sqrt{k}}{a}-\frac{ig}{N} (\lambda \dot{\bar{\lambda}}+\bar{\lambda} \dot{\lambda})+2(\psi \bar{\lambda}-\bar{\psi} \lambda) \frac{\sqrt{k}}{a} g+\lambda \bar{\lambda} \frac{\sqrt{k}}{a} g-2\psi \bar{\psi} \lambda \bar{\lambda} g+\frac{i}{N} (\dot{\lambda} \bar{\gamma}+\dot{\bar{\lambda}} \gamma)  \nonumber\\
	-\eta \bar{\gamma}+\bar{\eta} \gamma+2i \frac{\dot{a}}{aN} (\lambda \bar{\gamma}+\bar{\lambda} \gamma)-i \frac{\dot{a}}{aN} (\psi \bar{\gamma}+\bar{\psi} \gamma)+(\psi \bar{\gamma}-\bar{\psi} \gamma) \frac{\sqrt{k}}{a}-\psi \bar{\psi} (\lambda \bar{\gamma}-\bar{\lambda} \gamma) \nonumber \\
	+3F (\psi \bar{\lambda}-\bar{\psi} \lambda)-3s g (\psi \bar{\lambda}-\bar{\psi} \lambda)+\frac{3}{2}g (\bar{\lambda} \eta-\lambda \bar{\eta})-2s (\lambda \bar{\gamma}-\bar{\lambda} \gamma)+\frac{\sqrt{k}}{a} (\lambda \bar{\gamma}-\bar{\lambda} \gamma) \nonumber\\
	+3i \frac{\dot{s}}{N} (\lambda \bar{\eta}+\bar{\lambda} \eta)+(F'-g) (\psi \bar{\eta}-\bar{\psi} \eta)+(\bar{\psi} \gamma-\psi \bar{\gamma}) s+3F \lambda \bar{\lambda}-3\eta \bar{\eta} s\Big)\Big]
\end{align}
On-shell 
\begin{align}\label{onshell}
	g=-\frac{1}{3} R+F'-4s^2+6 \frac{\sqrt{k}}{a} s+...
\end{align}
where $...$ indicates terms containing fermions.

\subsubsection{Starobinsky-massive scalar field}
Now we specialize to the case $F(s)=\frac{4}{3} s^3$. A convenient change of variables is in order,
\begin{align}\label{n2trans}
	\phi=a(1-\alpha g), && \chi=-\alpha \gamma a+\lambda a(1-\alpha g), && \bar{\chi}=- \alpha \bar{\gamma} a+\bar{\lambda} a (1-\alpha g)
\end{align}
A further re-scaling of the fermions finally yields 
\begin{align*}
	L=\frac{3}{\kappa} N \Big[-a \frac{\dot{\phi} \dot{a}}{N^2}+k\phi-\frac{1}{4\alpha} a (\phi-a)^2+{a}^{3} \big(\alpha \frac{\dot{s}^{2}}{N^2}-{s}^{2}\big)-3 \sqrt{k} a s \big(\phi-a+\frac{4}{3}\alpha a {s}^{2}\big)+\frac{i}{N} \big(\lambda \dot{\bar{\chi}}+\bar{\lambda} \dot{\chi} \nonumber \\
	-(\lambda \bar{\chi}+\bar{\lambda} \chi) \frac{\dot{a}}{a}+(\psi \bar{\lambda}+\bar{\psi} \lambda) a \dot{\phi}+(\psi \bar{\chi}+\bar{\psi} \chi) \dot{a}\big)-3\lambda \bar{\lambda} a s+\frac{1}{2}\phi (\lambda \bar{\eta}- \bar{\lambda} \eta) {a}^{-\frac{1}{2}}-(\lambda \bar{\chi}-\bar{\lambda} \chi) \frac{\sqrt{k}}{a} \nonumber \\
	-\phi (\psi \bar{\lambda}-\bar{\psi} \lambda) \sqrt{k}+\phi \lambda \bar{\lambda} \frac{\sqrt{k}}{a}+(\eta \bar{\chi}-\bar{\eta} \chi) {a}^{-\frac{1}{2}}-\sqrt{k} (\psi \bar{\chi}-\bar{\psi} \chi)+\psi \bar{\psi} (\lambda \bar{\chi}-\bar{\lambda} \chi)+2s (\lambda \bar{\chi}-\bar{\lambda} \chi) \nonumber \\	
\end{align*}
\begin{align}\label{lagmassive}
		-3 (\psi \bar{\lambda}-\bar{\psi} \lambda) {a}^{2} s-(\psi \bar{\eta}-\bar{\psi} \eta) {a}^{\frac{3}{2}}+(\psi \bar{\chi}-\bar{\psi} \chi) a s+\phi (\psi \bar{\eta}-\bar{\psi} \eta) a^{\frac{1}{2}}+2\phi (\psi \bar{\lambda}-\bar{\psi} \lambda) a s \nonumber \\
	-\frac{3}{2} (\lambda \bar{\eta}-\bar{\lambda} \eta) a^{\frac{1}{2}}+\alpha \big(\frac{i}{N} \big(\dot{\bar{\eta}} \eta+\dot{\eta} \bar{\eta}+3 (\lambda \bar{\eta}+\bar{\lambda} \eta) a^{\frac{1}{2}} \dot{s}-2 (\psi \bar{\eta}+\bar{\psi} \eta) {a}^{\frac{3}{2}} \dot{s}\big)+6 (\lambda \bar{\eta}-\bar{\lambda} \eta) a^{\frac{1}{2}} {s}^{2} \nonumber \\
	-2\psi \bar{\psi} \eta \bar{\eta}+\frac{3}{2} a^{-2} \lambda \bar{\lambda} \eta \bar{\eta}+5\eta \bar{\eta} s+4 (\psi \bar{\eta}-\bar{\psi} \eta) {a}^{\frac{3}{2}} {s}^{2}+3\eta \bar{\eta} \frac{\sqrt{k}}{a}+4\lambda \bar{\lambda} a {s}^{3}+4 (\psi \bar{\lambda}-\bar{\psi} \lambda) {a}^{2} {s}^{3}\big)\Big]
\end{align}
Note that we are allowing for positive $k$.

Performing the Legendre transformation $H=\dot{a} p_a+\dot{\phi} p_\phi-\dot{\lambda} \pi_{\bar{\lambda}}+\dot{\bar{\lambda}} \pi_\lambda-\dot{\chi} \pi_{\bar{\chi}}+\dot{\bar{\chi}} \pi_\chi-\dot{\eta} \pi_{\bar{\eta}}+\dot{\bar{\eta}} \pi_\eta-L$ we get the Hamiltonian constraint
\begin{align}\label{quadraticsQa}
	H_0&=\frac{\kappa}{3} \big(-a^{-1} p_a p_\phi+\frac{1}{4\alpha} a^{-3} p_s^2\big)+V(a,\phi,s)-i (\lambda \bar{\chi}+\bar{\lambda} \chi) a^{-2} p_\phi-\frac{3}{2} (\lambda \bar{\eta}+\bar{\lambda} \eta) i p_s a^{-\frac{5}{2}} \nonumber\\
	&+\frac{\kappa}{3} \Big(3\lambda \bar{\lambda} a s-\frac{9}{2} \alpha a^{-2} \lambda \bar{\lambda} \eta \bar{\eta}-6 (\lambda \bar{\eta}-\bar{\lambda} \eta) {s}^{2} a^{\frac{1}{2}} \alpha-\frac{1}{2} \phi (\lambda \bar{\eta}-\bar{\lambda} \eta) {a}^{-\frac{1}{2}}-2 (\lambda \bar{\chi}-\bar{\lambda} \chi) s \nonumber\\
	&+\frac{3}{2} (\lambda \bar{\eta}-\bar{\lambda} \eta) a^{\frac{1}{2}}-\frac{3}{2}\alpha \lambda \bar{\lambda} \eta \bar{\eta} {a}^{-2}-3\eta \bar{\eta} \frac{\sqrt{k}}{a} \alpha-(\eta \bar{\chi}-\bar{\eta} \chi) {a}^{-\frac{1}{2}}-4\alpha \lambda \bar{\lambda} a {s}^{3}-\phi \lambda \bar{\lambda} \frac{\sqrt{k}}{a} \nonumber \\
	&+(\lambda \bar{\chi}-\bar{\lambda} \chi) \frac{\sqrt{k}}{a}-5\eta \bar{\eta} s \alpha\Big) \approx 0,
\end{align}
with the potential $V$ given by
\begin{align}\label{potstar}
	V(\phi,a,s)=\frac{3}{\kappa} \Big[-k \phi+\frac{1}{4\alpha} a (\phi-a)^2+a^3 {s}^{2}+3\sqrt{k} a s \big(\frac{4}{3} \alpha a s^2+\phi-a\big)\Big],
\end{align}
The supersymmetric constraints are 
\begin{align}\label{susycs}
	-S&=\lambda (i p_a-W_a)+\chi \frac{1}{a} (i p_\phi-W_\phi)+\eta a^{-\frac{3}{2}} (i p_s-W_s)-\frac{3}{\kappa a} \lambda (\bar{\lambda} \chi+3\alpha \eta \bar{\eta}) \approx 0,\\
	\bar{S}&=\bar{\lambda} (i p_a+W_a)+\bar{\chi} \frac{1}{a} (i p_\phi+W_\phi)+\bar{\eta} {a}^{-\frac{3}{2}} (i p_s+W_s)-\frac{3}{\kappa a} \bar{\lambda} (\lambda \bar{\chi}-3\alpha \eta \bar{\eta}) \approx 0,
\end{align}
with superpotential
\begin{align}\label{superpot}
	W(a,\phi,s)=\frac{3}{\kappa} \big(\sqrt{k} \phi a-a^2 s (\phi-a)-\frac{4}{3} \alpha {s}^{3} a^3\big),
\end{align}
which is related to the scalar potential (\ref{potstar}) according to
\begin{align}\label{potfromsuper}
V(a,\phi,s)=\frac{1}{2} G^{AB} W_A W_B=\frac{\kappa}{3} \big(-a^{-1} W_a W_\phi+\frac{1}{4 \alpha} a^{-3} W_s^2\big)	
\end{align}

On the other hand, the second-class constraints $\pi_\lambda\approx 0$, $C_2=\pi_{\bar \lambda}\approx 0$, $C_3=\pi_\chi+\frac{3i}{\kappa} \lambda\approx 0$, $C_4=\pi_{\bar \chi}-\frac{3i}{\kappa} \bar{\lambda}\approx 0$, $C_5=\pi_{\eta}-\frac{3i}{\kappa} \alpha \eta\approx 0$, and $C_6=\pi_{\bar \eta}+\frac{3i}{\kappa} \alpha \bar{\eta}\approx 0$ lead to the Dirac brackets $\{q_i,p_j\}_D=\delta_{ij}$, and
\begin{align}
	\{\lambda,\bar{\chi}\}_D=\frac{\kappa}{3i},  && \{\bar{\lambda},\chi\}_D=\frac{\kappa}{3i}, && \{\eta,\bar{\eta}\}_D=\frac{i \kappa}{6 \alpha}.
\end{align}
Note that $\{\lambda,\bar{\lambda}\}_D=0$, $\{\chi,\bar{\chi}\}_D=0$.

\subsubsection{Quantization}\label{qn2}
The complex fermions satisfy the non-vanishing anti-commutators
\begin{align}\label{n2algebra}
	[\lambda,\bar{\chi}]_+=\frac{\hbar \kappa}{3}=[\bar{\lambda},\chi]_+, && [\eta,\bar{\eta}]_+=-\frac{\hbar \kappa}{6 \alpha}.
\end{align}
Thus, we use operators $A, A^\dagger$ and $B, B^\dagger$ defined in (\ref{fdrladder}), as well as 
\begin{align}
C&=\Big(\frac{6\alpha}{\hbar \kappa}\Big)^{\frac{1}{2}} \eta, && C^\dagger=\Big(\frac{6 \alpha}{\hbar \kappa}\Big)^{\frac{1}{2}} \bar{\eta},
\end{align}	
so that we now have three fermionic oscillators,
\begin{align}
	[A,A^\dagger]_+=-1, && [B,B^\dagger]_+=1, && [C,C^\dagger]_+=-1.
\end{align}
The space of states is now constituted by the vacuum $\ket{1}$, together with $\ket{2}=A^\dagger \ket{1}$, $\ket{3}=B^\dagger \ket{1}$, $\ket{4}=C^\dagger \ket{1}$, $\ket{5}=A^\dagger B^\dagger \ket{1}$, $\ket{6}=A^\dagger C^\dagger \ket{1}$, $\ket{7}=B^\dagger C^\dagger \ket{1}$, $\ket{8}=A^\dagger B^\dagger C^\dagger \ket{1}$. They are orthogonal and their norm can be either positive or negative: $\braket{1|1}=-\braket{2|2}=\braket{3|3}=-\braket{4|4}=-\braket{5|5}=\braket{6|6}=-\braket{7|7}=\braket{8|8}$.

Choosing the vacuum of positive norm, the induced matrix representation of the creation/annihilation operators, as defined in Section \ref{secfdr}, is
\begin{subequations}\label{matrices}
	\begin{align}
		A&=\begin{pmatrix}
			\sigma_+ & 0 & 0 & 0\\
			0 & 0 & 1 & 0\\
			0 & 0 & 0 & 0\\
			0 & 0 & 0 & \sigma_+
		\end{pmatrix}, && A^\dagger=\begin{pmatrix}
			-\sigma_- & 0 & 0 & 0 \\
			0 & 0 & 0 & 0 \\
			0 & -1 & 0 & 0\\
			0 & 0 & 0 & -\sigma_-
		\end{pmatrix}, \\
		B=&\begin{pmatrix}
			0 & \zeta_+ & -\sigma_- & 0\\
			0 & 0 & 0 & \sigma_-\\
			0 & 0 & 0 & -\zeta_-\\
			0 & 0 & 0 & 0
		\end{pmatrix},
		&&
		B^\dagger=\begin{pmatrix}
			0 & 0 & 0 & 0\\
			\zeta_+ & 0 & 0 & 0\\
			-\sigma_+ & 0 & 0 & 0\\
			0 & \sigma_+ & -\zeta_- & 0
		\end{pmatrix},\\
		C&=\begin{pmatrix}
			0 & \sigma_+ & -\zeta_- & 0\\
			0 & 0 & 0 & -\zeta+ \\
			0 & 0 & 0 & \sigma_+\\
			0 & 0 & 0 & 0 
		\end{pmatrix}, && C^\dagger=\begin{pmatrix}
			0 & 0 & 0 & 0\\
			-\sigma_- & 0 & 0 & 0\\
			\zeta_- & 0 & 0 & 0\\
			0 & \zeta_+ & -\sigma_- &0   
		\end{pmatrix}.
	\end{align}	
\end{subequations}
where $\sigma_\pm=\frac{1}{2} (\sigma_1\pm i \sigma_2)$ and $\zeta_\pm=\frac{1}{2} (1\pm \sigma_3)$.

The quantum state of the universe has eight components
\begin{align}
	\Psi(\phi,a,s)=\begin{pmatrix}
		\psi_1(a,\phi,s) \\ \psi_2(a,\phi,s)\\ ... \\ \psi_8 (a,\phi,s)
	\end{pmatrix}
\end{align}
and is annihilated by the supercharge operators.

A feature of the supersymmetric theories is that the ordering problem is somewhat alleviated compared to the non-supersymmetric case, since there are fewer choices with a linear momentum: $i p_a \to (1-y) i p_a+y a^{-p} i p_a a^p=\hbar (\partial_a+\frac{p y}{a})$. The constant $y$ already accounts for different orderings of the cubic fermionic terms since $(\lambda \bar{\lambda} \chi+3\alpha \lambda \eta \bar{\eta})_{\text{classical}} \to (1+u) \lambda \bar{\lambda} \chi+u \lambda \chi \bar{\lambda}+3\alpha ((1+v) \lambda \eta \bar{\eta}+v \lambda \bar{\eta} \eta)=\lambda \bar{\lambda} \chi+3\alpha \lambda \eta \bar{\eta}-\lambda \frac{\hbar}{a} (u-\frac{3}{2} v)$, using (\ref{n2algebra}), and the combination of constants $u, v$ can be absorbed into $y$. Thus, we write the supercharges as follows,
\begin{align}
	-S&=\lambda \big(\hbar (\partial_a+\frac{y}{a})-W_a\big)+\chi \frac{1}{a} \big(\hbar (\partial_\phi+\frac{x}{\phi})-W_\phi \big)+\eta a^{-\frac{3}{2}} (\hbar \partial_s-W_s)-\frac{3}{\kappa a} \big(\lambda \bar{\lambda} \chi+3\alpha \lambda \eta \bar{\eta}\big) \\
	\bar{S}&=\bar{\lambda} \big(\hbar (\partial_a-\frac{y}{a})+W_a\big)+\bar{\chi} \frac{1}{a} (\hbar (\partial_\phi-\frac{x}{\phi})+W_\phi)+\bar{\eta} {a}^{-\frac{3}{2}} (\hbar \partial_s+W_s)+\frac{3}{\kappa a} \big(\bar{\chi} \lambda \bar{\lambda}+3 \alpha \eta \bar{\eta} \bar{\lambda}\big)=S^\dagger
\end{align}	
Note that a third parameter can be introduced due to $p_s$. 

The quantum supersymmetric constraints $S\ket{\Psi}=0=\bar{S}\ket{\Psi}$ yield fourteen first order PDE's for the eight components of the state. Now we proceed to investigate the physical content of these equations.

\subsubsection{Coupled equations and the classical limit}
From the fourteen equations, we get two independent sets of four equations for $\psi_2, \psi_3, \psi_4$, and $\psi_5, \psi_6, \psi_7$. We will only consider the first set as the second one yields similar expressions. Defining $A=\psi_2-\psi_3$, $B=\psi_2+\psi_3$, the first set can be expressed as 
\begin{subequations}\label{set1}
	\begin{align}
		(\hbar \partial_a+\frac{\hbar}{2 a}) B-\frac{1}{a} (\frac{\hbar x}{\phi}-W_\phi) A+\frac{1}{2 \sqrt{\alpha}} a^{-\frac{3}{2}} (\hbar \partial_s-W_s) \psi_4=0, \\
		\frac{\hbar}{a} \partial_\phi A-((1+y) \frac{\hbar}{a}-W_a) B-\frac{1}{2 \sqrt{\alpha}} a^{-\frac{3}{2}} (\hbar \partial_s-W_s) \psi_4=0, \\
		\frac{1}{a} (\hbar (\partial_\phi-\frac{x}{\phi})+W_\phi) \psi_4 +\frac{1}{2 \sqrt{\alpha}} {a}^{-\frac{3}{2}} (\hbar \partial_s+W_s) B=0, \\
		(\hbar (\partial_a-\frac{y}{a})+W_a) \psi_4-\frac{1}{2\sqrt{\alpha}} {a}^{-\frac{3}{2}} (\hbar \partial_s+W_s) A=0.
	\end{align}
\end{subequations}
with the partial derivatives of the superpotential (\ref{superpot}) given by 
\begin{subequations}\label{superderiv}
	\begin{align}
		W_a&=\frac{3}{\kappa}\big(\sqrt{k} \phi-2 a s \phi+3 a^2 s-4 \alpha {s}^{3} a^2 \big), \\
		W_\phi&=\frac{3}{\kappa} \big(\sqrt{k} a-a^2 s\big),\\
		W_s&=\frac{3}{\kappa} \big(-a^2 (\phi-a)-4 \alpha s^2 a^3\big)
	\end{align}
\end{subequations}
We look for solutions of the form $B(a,\phi,s)=\tilde{B}(a,\phi,s) \exp(\frac{i}{\hbar} S(a,\phi,s))$, $A(a,\phi,s)=\tilde{A}(a,\phi,s) \exp(\frac{i}{\hbar} S(a,\phi,s))$, $\psi_4(a,\phi,s)=\tilde{\psi}(a,\phi,s) \exp(\frac{i}{\hbar} S(a,\phi,s))$. Substituting these expressions into (\ref{set1}) and factorizing the exponential we get, to zeroth order in Planck's constant, 
\begin{subequations}\label{constwkb}
\begin{align}
	i \tilde{B} \partial_a S+\frac{1}{a} W_\phi \tilde{A}+\frac{1}{2 \sqrt{\alpha}} {a}^{-\frac{3}{2}} \tilde{\psi} (i \partial_s S-W_s)=0, \label{01}\\
	\frac{i}{a} \tilde{A} \partial_\phi S+W_a \tilde{B}-\frac{1}{2 \sqrt{\alpha}} {a}^{-\frac{3}{2}} \tilde{\psi} (i \partial_s S-W_s)=0, \label{02}\\
	\frac{1}{a} \tilde{\psi} (i \partial_\phi S+W_\phi)+\frac{1}{2 \sqrt{\alpha}}{a}^{-\frac{3}{2}} \tilde{B} (i \partial_s S+W_s)=0, \label{03}\\
	\tilde{\psi} (i \partial_a S+W_a)-\frac{1}{2 \sqrt{\alpha}}{a}^{-\frac{3}{2}} \tilde{A} (i \partial_s S+W_s)=0. \label{04}
\end{align}
\end{subequations}
Assuming $i \partial_s S+W_s\ne 0$ and $\tilde{\psi}\ne0$, we solve for $\frac{\tilde{B}}{\tilde{\psi}}$, $\frac{\tilde{A}}{\tilde{\psi}}$ from (\ref{03}) and (\ref{04}). Substituting them into (\ref{01}) and (\ref{02}), and using (\ref{potfromsuper}), we get two copies of the Hamilton-Jacobi (H-J) equation for the bosonic part of the Hamiltonian (\ref{quadraticsQa})
\begin{align}\label{hj}
	-\frac{1}{a} (\partial_\phi S) \partial_a S+\frac{1}{4 \alpha} {a}^{-3} (\partial_s S)^2+\frac{3}{\kappa} V(a,\phi,s)=0.
\end{align}
Setting $s=0$ and neglecting partial derivatives with respect to $s$, we obtain the H-J equation for the pure model of Starobinsky, whose solutions take part in the phase of the WKB wavefunction (\ref{wkbk1}).

On the other hand, at first order in Planck's constant, we get
\begin{subequations}\label{fi}
	\begin{align}
		(\partial_a \tilde{B}+\frac{1}{2a} \tilde{B})-\frac{x}{a \phi} \tilde{A}+\frac{1}{2 \sqrt{\alpha}} a^{-\frac{3}{2}} \partial_s \tilde{\psi}=0, \\
		\frac{1}{a} \partial_\phi \tilde{A}-\frac{1+y}{a} \tilde{B}-\frac{1}{2 \sqrt{\alpha}} a^{-\frac{3}{2}} \partial_s \tilde{\psi}=0,\\
		\partial_\phi \tilde{\psi}-\frac{x}{\phi} \tilde{\psi}+\frac{1}{2 \sqrt{\alpha}} a^{-\frac{3}{2}} \partial_s \tilde{B}=0, \\
		\partial_a \tilde{\psi}-\frac{y}{a} \tilde{\psi}-\frac{1}{2 \sqrt{\alpha}} a^{-\frac{3}{2}} \partial_s \tilde{A}=0.
	\end{align}
\end{subequations}

Furthermore, given a solution $S(a,\phi,s)$, equations (\ref{constwkb}) give further relations among the coefficients $\tilde{A}, \tilde{B}, \tilde{\psi}$. For example, taking $S=i \frac{1}{\kappa \sqrt{\alpha}} a^{\frac{3}{2}} \phi^{\frac{3}{2}}$, valid for $k=0$, 
\begin{align}\label{zeroconst}
	\tilde{A}=-\phi \tilde{B}, && \tilde{\psi}=\mp i \phi^{\frac{1}{2}} \tilde{B}
\end{align}
Substituting them into (\ref{fi}) we get
\begin{subequations}\label{leq}
	\begin{align}
		(\partial_a \tilde{B}+\frac{\tilde{B}}{2a})+\frac{x}{a} \tilde{B}=0, \label{l1}\\
		\partial_\phi \tilde{B}+(2+y) \frac{1}{\phi} \tilde{B}=0, \label{l2}\\
		\partial_\phi \tilde{B}+(\frac{1}{2}-x) \frac{1}{\phi} \tilde{B}=0, \label{l3}\\
		\partial_a \tilde{B}-\frac{y}{a} \tilde{B}=0. \label{l4}
	\end{align}
\end{subequations}
We are getting too many equations because we are completely neglecting any dependence on $s$, but the bosonic minisuperspace is really tridimensional. Nonetheless, for appropriate values of $x$ and $y$, we can make equations (\ref{leq}) hold approximately. Indeed, (\ref{l4}) and (\ref{l3}) yield $\tilde{B}=a^y \phi^{x-\frac{1}{2}}$. If, for example, we set $y=2$ and $x=-\frac{1}{2}$, then (\ref{l1}) and (\ref{l2}) are proportional to $\frac{a}{\phi}\approx 0$ and $(\frac{a}{\phi})^2\approx 0$, respectively.

\subsubsection{Decoupled equations}
On the other hand, $\psi_1$ and $\psi_8$ satisfy each a set of three equations that, choosing $y=-\frac{1}{4}$, $x=0$, read
\begin{subequations}\label{foe}
\begin{align}
	(\partial_a-\frac{5}{4a}+\frac{1}{\hbar} W_a) \psi_1=0, && (\partial_a-\frac{5}{4a}-\frac{1}{\hbar} W_a) \psi_8=0, \\
	(\partial_\phi+\frac{1}{\hbar} W_\phi) \psi_1=0, && (\partial_\phi-\frac{1}{\hbar} W_\phi) \psi_8=0, \\
	(\partial_s+\frac{1}{\hbar} W_s) \psi_1=0, && (\partial_s-\frac{1}{\hbar} W_s) \psi_8=0.
\end{align}
\end{subequations}
whose solutions are given by
\begin{subequations}
\begin{align}
	\psi_1(a,\phi,s)&=a^{\frac{5}{4}} \exp \big(-\frac{1}{\hbar} W(a,\phi,s)\big), \\
	\psi_8(a,\phi,s)&=a^{\frac{5}{4}} \exp \big(\frac{1}{\hbar} W(a,\phi,s)\big),
\end{align}
\end{subequations}
in terms of the superpotential (\ref{superpot}). Exponential wave functions are associated to pure quantum states with no classical analogue, since $W(a,\phi,s)$ solves the Hamilton-Jacobi equation (\ref{potstar}) for Euclidean geometry $p_A \to -i p_A$. In the limit $\alpha=0$, $\phi$ goes over to $a$ and we recover the celebrated no boundary ($e^{\frac{3}{\hbar \kappa} a^2}$) and wormhole ($e^{-\frac{3}{\hbar \kappa} a^2}$) states for closed universes $k=1$.

Setting $\psi_2=0=...=\psi_7$, and computing $[S,\bar{S}]_+ \Psi=0$, we find the second order equations satisfied by $\psi_1$, $\psi_8$,
\begin{subequations}\label{n2second}
\begin{align}
	\Big[\frac{1}{a} (\partial_\phi-\frac{1}{\hbar} W_\phi) (\partial_a-\frac{5}{4 a}+\frac{1}{\hbar} W_a)+(\partial_a-\frac{1}{4a}-\frac{1}{\hbar} W_a) \frac{1}{a} (\partial_\phi+\frac{1}{\hbar} W_\phi)& \nonumber \\
	-\frac{1}{2 \alpha} a^{-3} (\partial_s-\frac{1}{\hbar} W_s) (\partial_s+\frac{1}{\hbar} W_s)+\frac{3}{2a} \frac{1}{a} (\partial_\phi+\frac{1}{\hbar} W_\phi)\Big]& \psi_1=0, \label{psi1n2eq}\\
	\Big[\frac{1}{a} ( \partial_\phi+\frac{1}{\hbar} W_\phi)  (\partial_a-\frac{5}{4a}-\frac{1}{\hbar} W_a)+( \partial_a+\frac{1}{4a}+\frac{1}{\hbar} W_a) \frac{1}{a} ( \partial_\phi-\frac{1}{\hbar} W_\phi)& \nonumber \\
	-\frac{1}{2 \alpha} {a}^{-3} ( \partial_s+\frac{1}{\hbar} W_s) ( \partial_s-\frac{1}{\hbar} W_s)+\frac{1}{a} \frac{1}{a} (\partial_\phi-\frac{1}{\hbar} W_\phi)\Big]& \psi_8=0.
\end{align}
\end{subequations}
Clearly, (\ref{n2second}) hold automatically if (\ref{foe}) do. Substituting (\ref{superderiv}) and expanding one identifies the non-supersymmetric Starobinsky WDW equation (\ref{starwdw}) plus contributions from the massive scalar field and some extra terms proportional to $\hbar$.
\begin{subequations}\label{n2sol}
	\begin{align}
		\psi_1(a,\phi,s)&=a^{\frac{5}{4}} \exp\big[\frac{-3}{\hbar \kappa} \big(\sqrt{k} \phi a-a^2 s (\phi-a)-\frac{4}{3} \alpha {s}^{3} a^3\big)\big], \label{psi1}\\
		\psi_8(a,\phi,s)&=a^{\frac{5}{4}} \exp \big[\frac{3}{\hbar \kappa}\big(\sqrt{k} \phi a-a^2 s (\phi-a)-\frac{4}{3} \alpha {s}^{3} a^3\big)\big].
	\end{align}
\end{subequations}
Or, recalling $\phi=a (1+\frac{\alpha}{3} R)$, 
\begin{align}
	\psi_1 (a,R,s)=a^{\frac{5}{4}} \exp\big[-\frac{3}{\hbar \kappa} \big(\sqrt{k} a^2 (1+\frac{\alpha}{3} R)-\frac{\alpha}{3} a^3 s R-\frac{4}{3} \alpha a^3 {s}^{3}\big)\big]
\end{align}

\begin{figure}[h!]
	\centering
	\includegraphics[width=0.4\textwidth]{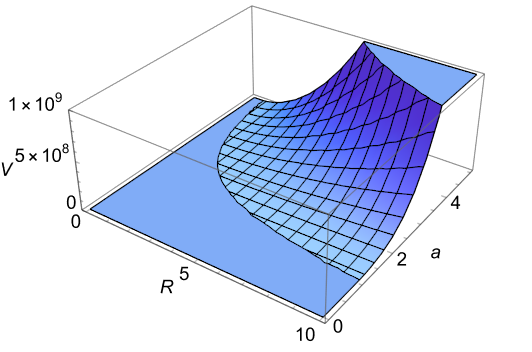}
	\caption{Scalar potential $U(a,R,s=0)$. The solid region indicates a negative value.}
	\label{Fig3}
\end{figure}

As an application of the exact solutions, we interpret the wave function as providing the probability distribution of initial values with which the universe began its classical evolution \cite{vilenkin88,mijic}. Thus, we evaluate the wave function near the Euclidean-Lorentzian boundary defined by the curve where the scalar potential (\ref{potstar0}) changes sign.  This occurs at $a \phi (1-\frac{a}{\phi})^2=4 \alpha$ ($k=1$), or 
\begin{align}\label{vanish}
	a^*(R)=2 \sqrt{\alpha} \frac{(1+\frac{\alpha}{3} R)^{\frac{1}{2}}}{\frac{\alpha}{3} R}=2 \sqrt{\alpha} \left( \frac{1}{\frac{\alpha}{3} R}+\frac{1}{(\frac{\alpha}{3} R)^2}\right)^{\frac{1}{2}}
\end{align}
which is depicted in Figure \ref{Fig3} (a). The value of $R$ along this path is interpreted as the initial value with which the universe started its classical evolution.

Since the exact wave functions are not normalizable in $s$, which is not unusual in quantum cosmology \cite{halliwell,wiltshire}, we use the projections of (\ref{n2sol}) on the plane $s=0$. Denoting $\psi_1(a,R,0)=\psi_-(a,R)$ and $\psi_8(a,R,0)=\psi_+(a,R)$, we have
\begin{align}\label{wave}
	\psi(a,R)_\pm=a^{\frac{5}{4}} \exp\big[\pm \frac{3}{\hbar \kappa} a^2 (1+\frac{\alpha}{3} R)\big]
\end{align}
Evaluating (\ref{wave}) at the curve (\ref{vanish}), 
\begin{align}
	\tilde{\psi}_\pm (R)=\psi_\pm (a^*,R)&=(\frac{1}{\frac{\alpha}{3} R}+\frac{1}{(\frac{\alpha}{3} R)^2})^{\frac{5}{8}} \exp \big[\pm \frac{3}{\hbar \kappa} \frac{4\alpha (1+\frac{\alpha}{3} R) (1+\frac{\alpha}{3} R)}{(\frac{\alpha}{3} R)^2}\big] \nonumber \\ 
	& \approx (\frac{1}{(\frac{\alpha}{3} R)})^{\frac{5}{8}} \exp \big[\pm \frac{12 \alpha}{\hbar \kappa} \big(1+\frac{2}{(\frac{\alpha}{3} R)}\big)\big] \propto \left(\frac{1}{R}\right)^{\frac{5}{8}} \exp \left( \pm \frac{72}{\hbar \kappa} \frac{1}{R} \right) , 
\end{align}
retaining up to first order terms in $(\frac{\alpha}{3} R)^{-1}$. Thus, setting $\kappa=8\pi G/2\pi^2$ and $\hbar=1$, we get 
\begin{align}\label{effective}
	\tilde{\psi}^2_\pm (R)=c^2 (\frac{1}{R})^{\frac{5}{4}} \exp\left( \pm \frac{36 \pi}{G} \frac{1}{R} \right) .
\end{align}
Note that it does not depend on $\alpha$. The functional dependence on $R$ of these probability distributions are roughly those found for the Vilenkin ($-$) and Hartle-Hawking ($+$) wave functions \cite{mijic,Henk1994}, although we have obtained them using real exponential wave functions of provided by quantum supersymmetric cosmology.  

On a constant $s$ surface, the mini superspace metric (see (\ref{potfromsuper})) $G_{AB}(a,R)$ is: $G_{aa}=-\frac{6}{\kappa} a (1+\frac{\alpha}{3} R)$, $G_{aR}=-\frac{\alpha}{\kappa} a^2=G_{R a}$, $G_{RR}=0$. Therefore, along $a^*(R)$,
\begin{align}
	ds^2=G_{aa} (da)^2+2 G_{aR} da \, dR=dR^2 \Big(G_{aa} \big(\frac{da^*}{dR}\big)^2+2 G_{aR} \frac{da}{dR}\Big) \approx 2 G_{aR} \frac{da^*}{dR} dR^2 \propto R^{-\frac{5}{2}} dR^2
\end{align}

Taking into account that above the Planck scale simple minisuperspace models are hardly viable, one computes the relative probability for the initial curvature to lie in the range $R_{\text{inf}}\le R_{i}\le R_{\text{cut}}$,  where (in Planck units $l_p=\sqrt{G}$) $R_{\text{min}} \approx 10^{-9} l_p^{-2}$ and $R_{\text{cut}}=l_p^{-2}$ given that $R_{\text{min}}\le R_i \le R_{\text{cut}}$, where $R_{\text{min}}\approx 10^{-11} G^{-1}$ \cite{mijic}, that is
\begin{align}
	P_\pm=\frac{\int_{R_{\text{inf}}}^{R_{\text{cut}}} R^{-\frac{5}{4}} \tilde{\psi}^2_\pm(R) dR}{\int_{R_{\text{min}}}^{R_{\text{cut}}} R^{-\frac{5}{4}} \tilde{\psi}^2_\pm(R) dR}=\frac{\int_{R_{\text{inf}}}^{R_{\text{cut}}} R^{-\frac{5}{2}} \exp\big(\pm \frac{36 \pi}{G} \frac{1}{R} \big) dR}{\int_{R_{\text{min}}}^{R_{\text{cut}}} R^{-\frac{5}{2}} \exp\big(\pm \frac{36 \pi}{G} \frac{1}{R} \big) dR}
\end{align}
A quick evaluation using Mathematica returns $P_-\approx 1-2.05\times 10^{-4.9\times 10^{10}}$ and $P_+\approx 4.76\times 10^{-4.86 \times 10^{12}}$, respectively. Thus, $\psi_-$ corresponds to an early universe with the initial conditions favoring the triggering of an inflationary phase. On the other hand, in the universe associated to $\psi_+$, inflation has a extremely small chance of occurring since it favors rather small values of curvature.

We remark, however, that the wave functions considered here are of the exponential type and there is no semi-classical behavior once the Universe tunnels to a configuration above the $a^*(R)$. For that to be the case, one would require a sort or supersymmetric breaking with the effect of transforming from exponential to oscillatory wave function. That is, (\ref{psi1n2eq}) holds and allows oscillatory wave functions even if (\ref{foe}) do not hold \cite{moniz2}. In fact, there are models of supersymmetric quantum cosmology where the only solution of the supersymmetric (and Lorentz) constraints is the trivial one $\Psi=0$ \cite{Cheng1995,moniz96}. This, however, does not prevent the existence of nontrivial solutions to the second order Hamiltonian constraint.

\subsection{Arbitrary potential}
Finally, we show the quantization with arbitrary $F$. In this case, the transformation (\ref{n2trans}) is not very useful, thus we stay with the original variables. The Hamiltonian is
\begin{align}\label{quadratics}
	H_0&=\frac{\kappa}{3} \Big(\frac{p_a p_g}{\alpha a^2}+\frac{b  p_g^{2}}{\alpha^2 a^3}+\frac{p_s^{2}}{4\alpha a^3}\Big)+V(a,g,s)-\frac{i \alpha p_a}{2 b a^2} (\lambda \bar{\gamma}+\bar{\lambda} \gamma)-\frac{i p_g}{2 a^3} (\lambda \bar{\gamma}+\bar{\lambda} \gamma)-\frac{i p_g}{a^3} (\lambda \bar{\gamma}+\bar{\lambda} \gamma) \nonumber \\
	&+\frac{3 i \alpha p_s}{4 a^3 \sqrt{b}} (\eta \bar{\gamma}+\bar{\eta} \gamma)-\frac{3i p_s}{2 a^3 \sqrt{b}} (\lambda \bar{\eta}+\bar{\lambda} \eta)
	+\frac{3}{\kappa} \Big(\frac{\sqrt{k}}{a} \lambda \bar{\lambda}-\frac{\alpha  \lambda \bar{\lambda}}{a^3 b} \big(\alpha \gamma \bar{\gamma}+6\eta \bar{\eta}\big)-\frac{3 \alpha^3}{2a^3 b}\eta \bar{\eta} \gamma \bar{\gamma}-s \lambda \bar{\lambda} \nonumber\\
	&-\frac{\alpha \sqrt{k}}{2a} (\lambda \bar{\gamma}-\bar{\lambda} \gamma)-\frac{3\alpha^{2} \sqrt{k}}{4 a}\gamma \bar{\gamma}+\frac{3}{2} \alpha s (\lambda \bar{\gamma}-\bar{\lambda} \gamma)+3\alpha \frac{s g}{b} \lambda \bar{\lambda}-3\alpha \frac{\sqrt{k}}{a} \eta \bar{\eta}+\frac{3 \alpha^{3}}{4} \frac{s g}{b} \gamma \bar{\gamma}-\alpha F'' \eta \bar{\eta} \nonumber\\
	&-\frac{3 \alpha F'}{2 \sqrt{b}} (\lambda \bar{\eta}-\bar{\lambda} \eta)+\alpha b^{\frac{1}{2}} (\eta \bar{\gamma}-\bar{\eta} \gamma)-3 \frac{\alpha F}{b} \lambda \bar{\lambda}-\frac{3 F'}{4 \sqrt{b}}  \alpha^{2} (\eta \bar{\gamma}-\bar{\eta} \gamma)+\frac{3}{2} \frac{\alpha g}{\sqrt{b}} (\lambda \bar{\eta}-\bar{\lambda} \eta)+3 \alpha a \eta \bar{\eta} \nonumber\\
	&+\frac{7}{4} \alpha^{2} s \gamma \bar{\gamma}+\frac{3 \alpha^{2} g}{4 \sqrt{b}} (\eta \bar{\gamma}-\bar{\eta} \gamma)+\frac{3 \alpha^2}{2 b} (\lambda \bar{\gamma}-\bar{\lambda} \gamma) (s g-F)-\frac{3\alpha^{3} F}{4 b} \gamma \bar{\gamma}-\frac{3\alpha^2}{a^3 b} \eta \bar{\eta} (\lambda \bar{\gamma}-\bar{\lambda} \gamma)\Big) \approx 0,
\end{align}
while the supersymmetric constraints are
\begin{align}
	-S&=\frac{\lambda}{\sqrt{ab}} (i p_a-W_a)+\frac{\gamma \sqrt{b}}{a^{3/2}} (i p_g-W_g)+\frac{\eta}{a^{3/2}} (i p_s-W_s)+\frac{3 \alpha \lambda}{2 \kappa a\sqrt{ab}} (\bar{\lambda} \gamma-\alpha \gamma \bar{\gamma}-6 \eta \bar{\eta})\approx 0, \label{sgen1}\\
	\bar{S}&=\frac{\bar{\lambda}}{\sqrt{ab}} (i p_a+W_a)+\frac{\bar{\gamma} \sqrt{b}}{a^{3/2}} (i p_g+W_g)+\frac{\bar{\eta}}{a^{3/2}} (i p_s+W_s)+\frac{3 \alpha \bar{\lambda}}{2\kappa a \sqrt{ab}} (\lambda \bar{\gamma}+\alpha \gamma \bar{\gamma}+6\eta \bar{\eta})\approx 0, \label{sgen2}
\end{align}
where $b$ stands for $1-\alpha g$ and $F=F(s)$.

The superpotential and scalar potential are, respectively,
\begin{align}
	W(a,g,s)&=\frac{3}{\kappa} \big[\sqrt{k} a^2 (1-\alpha g)+\alpha a^{3} (s g-F)\big]\\
	V(a,g,s)&=\frac{3}{\kappa} \big[-k a b+a^3 {s}^{2}+\frac{\alpha}{4} a^3 (F'-g)^2+\alpha a^3 (2s^2 g-3sF)+3\alpha \sqrt{k} a^2 (F-s g)\big] \label{e2}
\end{align}
The second class-constraints are: $\pi_\lambda+\frac{3i}{\kappa} \lambda \approx 0$, $\pi_{\bar{\lambda}}-\frac{3i}{\kappa} \bar{\lambda}\approx 0$, $\pi_\gamma-\frac{3i}{\kappa} \alpha \lambda\approx 0$, $\pi_{\bar{\gamma}}+\frac{3i}{\kappa} \alpha \bar{\lambda}\approx 0$, $\pi_\eta-\frac{3i}{\kappa} \alpha \eta\approx 0$, $\pi_{\bar{\eta}}+\frac{3i}{\kappa} \alpha \bar{\eta}\approx 0$, leading to the Dirac brackets $\{q_i,p_j\}_D=\delta_{ij}$ and $\{\lambda,\bar{\gamma}\}_D=\frac{i \kappa}{3 \alpha}$,  $\{\bar{\lambda},\gamma\}_D=\frac{i \kappa}{3 \alpha}$, $\{\gamma,\bar{\gamma}\}_D=\frac{2 i \kappa}{3 \alpha^2}$, $\{\eta,\bar{\eta}\}_D=\frac{i \kappa}{6 \alpha}$.

Using Weyl ordering for the supersymmetric constrains (\ref{sgen1}), (\ref{sgen2}), we obtain the exact solutions
\begin{align}
	\psi_\pm (a,g,s)=a^{\frac{7}{4}} \exp \left[ \pm \frac{3}{\hbar \kappa} \left( \alpha a^3 (F-g s)-\sqrt{k} a^2 (1-\alpha g)\right) \right] 
\end{align}
The power $\frac{7}{4}$ results from the different ordering compared to  (\ref{n2sol}). Making a transformation $g \to R$ dictated by the classical relation (\ref{onshell}), we get 
\begin{align}
	\psi(a,g,s)=a^{\frac{7}{4}} \exp \left\lbrace \pm \frac{3}{\hbar \kappa} \Big[\alpha a^3 \big(F+\frac{1}{3} s R+4 s^3-s F'-6\frac{\sqrt{k}}{a} s^2\big) -\sqrt{k} a^2 \big(1+\frac{\alpha}{3}R+4 \alpha s^2-6 \alpha \frac{\sqrt{k}}{a} s-\alpha F')\big) \Big] \right\rbrace 
\end{align}
For $k=0$, note that choosing $F=2s^3$ instead of $\frac{4}{3} s^3$ leaves the exponent $W\propto a^3 s R$. If we make a shift $F\to 2s^3+\frac{1}{\alpha} F(s)$, we get
\begin{align}
	\psi_\pm(a,g,s)=a^{\frac{7}{4}} \exp \left[ \pm \frac{3}{\hbar \kappa} \left( \frac{\alpha}{3} a^3  s R+a^3  (F-s F')\right) \right] 
\end{align}
Setting $\alpha=0$ in the last expression, one recovers the wave functions of the pure $F(\mathcal{R})$ action.

\section{Conclusion}\label{sec5}
We studied the quantum cosmology of two supersymmetric extensions of the FLRW model of Starobinsky and related higher derivative models. 

The supersymmetric N=1 and N=2 quadratic curvature models possess three fermionic degrees of freedom. Therefore, their second order formulation required the introduction of two dynamical superfields. In contrast, for the $F(\mathcal{R})$ action, with two fermionic degrees of freedom, only one extra superfield was needed. With the classically equivalent actions at hand, we found that, as with ordinary theories, the supersymmetric constraints are naturally given in terms of a superpotential. It would be interesting to derive the Hamiltonian and supercharge operators with the procedure described in  \cite{Lidsey2000}  and the superpotentials found here.

For quantization, the fermionic sector must be represented in some way. We used matrix representations acting on quantum states given by column vectors of dimension $2^{\lfloor n_f \rfloor}$ where $n_f$ is the number of physical fermions ($n_f=\frac{3}{2}$ for the N=1 model, $n_f=4$ for N=2 $F(\mathcal{R})$ and $n_f=3$ for the N=2 $(\nabla \mathcal{R})^2+...$) and ${\lfloor ... \rfloor}$ stands for the floor function. Furthermore, the components are wave functions defined over the bosonic minisuperspace.

Following Dirac quantization, the constraint operators annihilate physical quantum states. A remarkable consequence of supersymmetry is the set of additional constraints and the associated algebra. The ordering ambiguity is partially reduced, since the momentum operators appear linearly in the supercharges versus quadratically in the Hamiltonian constraint.

Substituting the WKB anzats $\psi(q_A)=G(q_A) e^{\pm \frac{i}{\hbar} S(q_A)}$, we get, at leading order, the Hamilton-Jacobi equation of the corresponding non-supersymmetric models. Moreover, given the solution $S$, the zeroth order equations impose further relations among the quantum prefactors of the different components $\psi_i$. In the N=1 case we obtain approximate solutions of the supersymmetric constraints. In the N=2 case, we obtained approximate solutions to the intermediate components, and exponential solutions of the form $e^{\pm W/\hbar}$ for the empty and filled components. The latter are associated with tunneling amplitudes in the classically forbidden region of mini-superspace. In fact, they are related to Euclidean solutions of the classical field equations, but in this context, they are defined all over the mini-superspace, regardless of the sign of the potential. 

While exponential wave functions are not adequate for the late universe, they might be appropriate for the very early universe. Thus, as an application we used them to compute the probability distribution of initial conditions. More precisely, we computed the probability density of the 4-curvature at the classical-quantum boundary, that is, where the potential ($k=1$) changes sign. Since these wave functions are not normalizable, we use their projection onto the plane of vanishing scalar field. These distributions resemble those obtained with the no-boundary/tunneling wave functions and the non-supersymmetric Starobinsky model. We computed the relative probability for the ``initial" curvature to be above the minimum value required for an appropriate inflationary phase. The returned probabilities are extremely close to 1 and 0, which are interpreted as universes where inflation occurs is a highly probable versus an extremely rare event. Also, exact exponential solutions were obtained for the N=2 models with an arbitrary superpotential. They will be investigated in the context of the problem of time \cite{ramirez2022} in future work.

There are still several uncovered aspects that will be addressed in subsequent work. For example, to detail a mechanism of supersymmetry breaking, making a component wave function transition from exponential to oscillatory behavior with classical limit. How to obtain an initial curvature distribution for $k=0$, when the potential has the same sign all over the minisuperspace.

On the other hand, the consistent results we have obtained in this and previous works using 1D supergravity encourage to look for a rigorous connection with other approaches to supersymmetric cosmology, in particular the dimensional reduction of 4D supergravity. This could shed light on the nature of the scalar fermions used in our approach and is necessary to introduce perturbations. Furthermore, it would be very interesting to investigate what our superfield method has to say about some intriguing results obtained using other approaches to FLRW supergravity \cite{Cheng1995,moniz96}.

\ \ \\

\centerline{\bf Acknowledgements}
We would like to thank the anonymous referee for helpful comments. We thank VIEP-BUAP and CONAHCyT for financial support. Some computations were done using Cadabra symbolic computer algebra system.

\bibliography{bibliography}
\end{document}